\documentclass[aip,amsmath,amssymb,reprint]{revtex4-1}

\usepackage{graphicx}
\usepackage{dcolumn}
\usepackage{bm}

\usepackage[utf8]{inputenc}
\usepackage[T1]{fontenc}
\usepackage{mathptmx}
\usepackage{etoolbox}

\usepackage{booktabs}
\usepackage{multirow}
\usepackage{float} 
\usepackage{amssymb}
\usepackage{amsmath}
\usepackage{xcolor}
\usepackage{longtable}
\usepackage{comment}
\usepackage{upgreek}

\makeatletter
\def\@email#1#2{%
 \endgroup
 \patchcmd{\titleblock@produce}
  {\frontmatter@RRAPformat}
  {\frontmatter@RRAPformat{\produce@RRAP{*#1\href{mailto:#2}{#2}}}\frontmatter@RRAPformat}
  {}{}
}%
\makeatother
\begin{document}

\preprint{AIP/123-QED}

\title[Analytical model for balling defects in laser melting using rivulet theory and solidification]{Analytical model for balling defects in laser melting using rivulet theory and solidification}

\author{Z. Taylor}
\affiliation{Department of Materials Science and Engineering, Stanford University}
\affiliation{SLAC National Accelerator Laboratory}
\author{T. Reddy}
\affiliation{Department of Materials Science and Engineering, Stanford University}
\affiliation{SLAC National Accelerator Laboratory}
\author{M. Fitzpatrick}
\affiliation{European Synchrotron Radiation Facility}
\author{K. Kim}
\affiliation{Department of Mechanical Engineering, University College London}
\affiliation{Research Complex at Harwell, Rutherford Appleton Laboratory}
\author{W. Li}
\affiliation{Department of Mechanical Engineering, University College London}
\affiliation{Research Complex at Harwell, Rutherford Appleton Laboratory}
\author{C.L.A. Leung}
\affiliation{Department of Mechanical Engineering, University College London}
\affiliation{Research Complex at Harwell, Rutherford Appleton Laboratory}
\author{P. D. Lee}
\affiliation{Department of Mechanical Engineering, University College London}
\affiliation{Research Complex at Harwell, Rutherford Appleton Laboratory}
\author{K. M. Bertsch}
\affiliation{Lawrence Livermore National Laboratory}
\author{L. Dresselhaus-Marais}
\affiliation{Department of Materials Science and Engineering, Stanford University}
\affiliation{SLAC National Accelerator Laboratory}
\email{leoradm@stanford.edu}

\date{\today}

\begin{abstract}
In the laser welding and additive manufacturing (AM) communities, the balling defect is primarily attributed to the action of fluid instabilities with a few authors suggesting other mechanisms. Without commenting on the validity of the fluid instability driven \textit{mechanism} of balling in AM, this work intends to present the most realistic analytical discussion of the balling defect driven purely by fluid instabilities. Synchrotron-based X-ray radiography of thin samples indicate that fluid instability growth rates and solidification can be comparable in magnitude and thus compete. Neglecting the action of fluid flows and heat transport, this work presents an analytical formalism which accounts for fluid instabilities and solidification competition, giving a continuous transition from balling to non-balling which is lacking in current literature. We adapt a Rivulet instability model from the fluid physics community to account for the stabilizing effects of the substrate which the Plateau-Rayleigh instability model does not account for, and estimate the instability growth rate. Our model predicts instability growth at higher wavelengths and shallower melt pool depths relative to width, as well as strong sensitivity to the solidification front curvature. Deviations between model predictions and our experimental results demonstrate the importance of fluid flows and heat transport in the balling process. Our experiments further demonstrate at least one mechanism by which the melt pool length and balling wavelength are not equivalent, as commonly claimed.
\end{abstract}

\maketitle

\section{\label{sec:intro}Introduction}

Laser powder bed fusion (LPBF) additive manufacturing (AM) has gained traction fabricating metallic components for numerous industries -- enabled by extensive research and development over the past two decades \cite{BlakeyMilner2021, Gradl2023}.
The productivity of the LPBF process today is fundamentally dependent on both laser power and scan speeds used to print parts. For faster prints, one can increase the laser power to melt/solidify a greater volume of material, and one can increase the scan speed to enable more material to be processed in the same amount of time. Simultaneously increasing both power and velocity would therefore be ideal to achieve the highest possible build rates. Despite the availability of high-power lasers and high-velocity laser scanning (galvo) mirrors, productivity rates today are restricted by the formation of balling defects at high laser power and scan velocity conditions. The deteriorated properties resulting from balling and the associated porosity they generate currently limit productivity. 

Numerous studies have investigated the effects of process parameters on keyhole, lack of fusion, and balling defects in various material systems with the goal of fabricating defect-free parts \cite{Gong2014, Cunningham2017, Johnson2019}. While the formation mechanisms of keyhole \cite{Cunningham2019, Zhao2020}, spatter \cite{Ly2017, Bidare2018}, and lack of fusion \cite{Tang2017, Snow2023} defects have been studied in detail, the mechanism resulting in balling formation remains poorly understood. The balling defect is characterized by the non-uniform accumulation of material along the length of the melt track, with periodic and aperiodic occurrences of elevated regions and discontinuities that form ``hills'' along the melt track \cite{Wang2021, Boutaous2021}. Consequently, these uneven melt tracks lead to non-uniform powder spreading, generating lack-of-fusion defects that are detrimental to the mechanical properties of the final part \cite{Sinclair2020, Fleming2023}. 
Bradstreet \cite{Bradstreet1968} was the first to report on balling, referred to as humping in the welding community, during gas metal arc welding. He observed that fluid non-uniformly accumulated at specific locations along the melt track. He attributed the growth of these accumulations to variations in the internal pressure driving the fluid flow through fluid channels connecting the melt pool and ball. Furthermore, Bradstreet hypothesized that the periodicity observed along the melt track could be explained by the Plateau-Rayleigh instability defined in fluid physics \cite{Bradstreet1968}.

The Plateau-Rayleigh instability describes the unstable growth of sinusoidal waves in a free-standing cylinder of fluid when subjected to any infinitesimal perturbation, in the absence of gravity or support surfaces \cite{Rayleigh_1878, Eggers_2008, mit}. 
Given sufficient time, these perturbations cause the cylindrical fluid jet to break lengthwise into discrete droplets, and the distance between droplets is defined as the instability wavelength, $\lambda$. The Plateau-Rayleigh analytical model gives a threshold that all wavelengths $\lambda/W>\pi$ are unstable and their amplitude will grow exponentially, for cylinder width $W=2R_0$ and radius $R_0$. For wavelengths longer than this threshold, the perturbation growth decreases the free energy of the surface causing it to grow over time and periodically fragment the fluid jet into droplets. Conversely, wavelengths shorter than this threshold increase the surface area and will not grow with time. An approximate correlation was observed between the analytically estimated wavelength from the Plateau-Rayleigh instability model and the spacing between balled regions in laser melting processes \cite{Bradstreet1968}. This led researchers to adapt Plateau-Rayleigh theory to laser-melting processes in order to define the threshold in terms of melt pool length ($L$) and width ($W$, equivalent to fluid width) \cite{Gratzke1992, Yadroitsev2010}. 

The initial threshold for balling in LPBF was a direct adaptation of the original Plateau-Rayleigh threshold, where $\lambda$ was assumed to be equivalent to $L$, setting the criterion as \(L/W > \pi\), which would result in balling. Gratzke et al.~\cite{Gratzke1992} and Yadroitsev et al.~\cite{Yadroitsev2010} made modifications to the Plateau-Rayleigh analytical model to account for aspects of melt pool geometry. These modifications assume a fluid cylinder intersecting the substrate and calculate the wavelength of a perturbation that would result in no net change in fluid surface area. This approach defines how the critical $L/W$ threshold changes based on the melt pool cylinder's position relative to the substrate. Gratzke et al.~\cite{Gratzke1992} proposed a threshold of \(L/W > 2\pi\), while Yadroitsev et al.~\cite{Yadroitsev2010} proposed a threshold of \(L/W > \frac{3}{2}\pi\). Both models are fundamentally similar, with the Gratzke model being a first-order Taylor expansion of the Yadroitsev criterion, and both result in comparable behavior. These analytical models neglect the effects of higher surface tension in metals \cite{Egry2010, Zhao2017}, fluid flows \cite{Hojjatzadeh2019, Guo2020}, and solidification \cite{Hooper2018, Mohammadpour2020} that are known to occur on fast timescales relevant to laser-melting processes. As a result, these threshold criterion cannot accurately predict balling, even when experimental $L$ and $W$ values are provided \cite{Francis, Leung2022, Bhatt2023}. Additionally, while a single threshold value suggests that balling occurs beyond a specific $L/W$ ratio, extensive experimental evidence shows that the transition to balling happens gradually, not abruptly \cite{Francis, Leung2022, Bhatt2023, Kurian}. The previous studies and inconsistencies further emphasize that the balling mechanism involves multiple competing phenomena. Furthermore, in molten metals, the effects of fluid instabilities cannot be isolated because they are continuously competing with the rapid time scales of solidification. Therefore, processing parameters that generate balling in a material system are identified through extensive single-track experiments during process optimization \cite{Johnson2019, Scime2019, Seede2020, Seede2021}.

While much of the literature assumes a Plateau-Rayleigh model or adaptations thereof to account for balling phenomena, some studies have instead attributed these dynamics only to fluid flows. \textit{In-situ} high-speed optical imaging studies of the melt pool, observed from a top view, have identified fluid accumulation resulting from vapor jet and fluid flows in the laser melting process, rather than to Plateau-Rayleigh instabilities \cite{Fabbro2010, Berger2011, Ai_2018, Xue_2022}. These works predict that at moderately high scan speeds, for which a vapor depression is present, the fluid is pushed to the rear by the vapor jet recoil pressure.
After fluid accumulation is initiated, the humped structures are held in place by surface tension and the surrounding solid material \cite{Wu_2017, Otto_2016}. As the laser advances, the hill volume continues to increase as more fluid is displaced by the vapor depression and/or the melting front \cite{Berger2011, Otto_2016}. Recently, Li et al.~\cite{Li2024} used synchrotron X-ray radiography to directly observe these balling mechanisms and validated their observations through numerical simulations. 

At this time, experiments and theory to describe balling typically attribute the phenomena to fluid instabilities like Plateau-Rayleigh or the dynamics of fluid flows, without accounting for the interplay between these physics and solidification. While not explored at this time, the multifaceted results from balling studies illustrate that a single mechanism does not cause balling, but results from the competition between vapor depression, fluid flows, solidification, and fluid instabilities. This interplay between driving forces ultimately determines the timescale over which balling occurs and the extent to which it can progress. To develop effective mitigation strategies, it is essential to study the interaction and competition between these mechanisms quantitatively.

In this paper, we present X-ray radiography experiments with an associated analytical framework for directly comparing the competition between fluid instabilities and solidification specifically. The impact of our work is the introduction of this framework, and future efforts could substitute more precise physical models of each component of the model. Our experimental results show that solidification rates and fluid instability rates can be of comparable magnitudes, emphasizing the need to develop more comprehensive models rather than relying solely on instability theories. We apply Rivulet instability theory and a simple solidification model to derive an analytical expression that describes the competition between the two and the balling transition, based on the attributes of the melt pool geometry.

We format this paper by first presenting our \textit{in-situ} radiography methods and findings (Sections \ref{sec:methods} and \ref{sec:experimental-results}), which are the inspiration for the analytical framework we introduce. We then present a more formal introduction of Plateau-Rayleigh instability theory and its assumptions before introducing the more physically grounded fluid instability of rivulets developed by the fluid instability community in Section \ref{sec:instability-theory}. With this groundwork introduced, we go on to adapt the rivulet instability model to additive manufacturing conditions to describe the rate of instability growth, before presenting our framework for the competition of fluid instabilities with solidification based on their relative growth rates in Section \ref{sec:model-development}. We then compare the predictions of our analytical framework for the specific case of our experimental results (Section \ref{sec:theory-to-experiments}) and balling in AM more generally (Section \ref{sec:implications}), before rigorously defining the advantages and limitations of our model in Section \ref{sec:limitations}.

\section{\label{sec:methods}X-ray radiography}

\begin{figure*}
    \centering
    \includegraphics[width=0.9\linewidth]{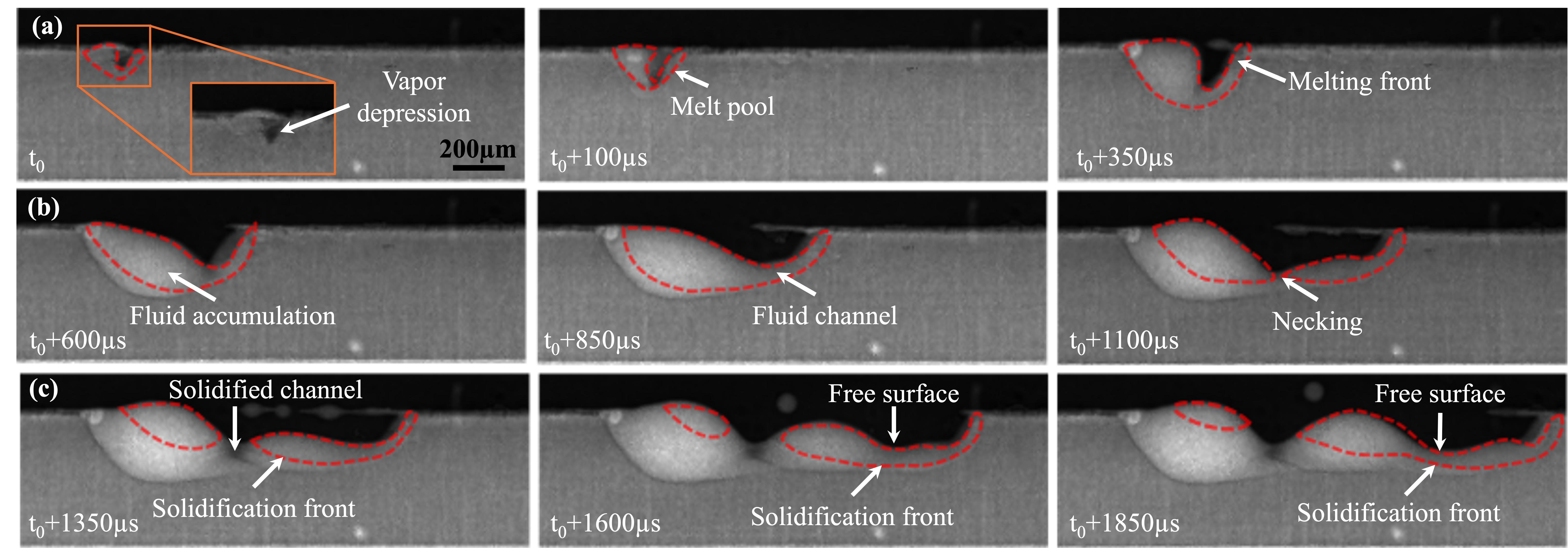}
    \caption{Selected X-ray images from the Ta laser melting experiment at \textit{P} = 300 W and \textit{v} = 500 mm/s. \textit{In-situ} images show (a) the melting front at the onset of the laser scan, (b) the fluid accumulation from flows, and (c) the solidification ``pinching'' the melt pool into hills. Important features observed during the process are indicated with white labelled arrows, and the liquid metal is denoted by red dotted lines.}
    \label{fig:TimeSeries}
\end{figure*}

The X-ray radiography experiments on LPBF in this work were performed at the ID19 beamline of the European Synchrotron Radiation Facility (ESRF). The laser melting experiments were carried out using the Quad-laser \textit{in-situ} and \textit{operando} process replicator (Quad-ISOPR), that is equipped with 4 ytterbium fiber lasers (SPI Lasers Ltd, UK) operated in continuous-wave mode with a wavelength of 1070 nm and a maximum power of 500 W. Laser focusing and scanning at the sample surface are achieved with a RenAM 500Q X–Y galvanometer scanner (Renishaw plc., UK) coupled with an $f$-theta scan lens to focus its spot size to 80 $\upmu$m. The ID19 beamline uses the U32 undulator to generate a polychromatic X-ray beam with a peak at $\sim$30 keV and harmonics up to 90 keV (each with a bandwidth of $\sim$2\%), and a total photon flux of approximately $10^{15}$ photons per second.

The high-energy X-rays used in this work illuminated pure tantalum (Ta) samples at normal incidence in transmission geometry as the sample was laser melted perpendicular to the X-rays. The Ta sample thickness along the X-ray beam direction was selected to be 100 µm to maximize the transmitted intensity for radiographic imaging. A single-crystal LuAG:Ce scintillator was used for imaging with a 5$\times$ visible objective lens on a commercial Photron FastCam SA-Z high-speed camera operated at a 40 kHz frame rate. The total track length for the laser was set to 4 mm, fitting into the microscope's field of view of 4.4 $\times$ 2.2 mm$^2$. Table \ref{tab:p-v} lists the process parameters used for laser melting experiments and our corresponding nomenclature for each scan. 

\begin{table}
\caption{\label{tab:p-v}Process parameters for laser melting of Ta.}
\begin{ruledtabular}
\begin{tabular}{ccc}
Case&Power ($P$)&Scan Velocity ($v$)\\
\hline
A & 200 & 250\\
B & 300 & 750\\
C & 300 & 500\\
D & 400 & 500\\
\end{tabular}
\end{ruledtabular}
\end{table}

\section{\label{sec:experimental-results}Experimental Results}

Fig.~\ref{fig:TimeSeries} includes a series of representative images from our radiography study that demonstrate the balling formation mechanism as the laser beam is scanned across the sample. This discussion illuminates both the importance of surface energy minimization in the rounded morphology of balling, as well as the influence of solidification and fluid flows that surface tension alone cannot account for. As shown in Fig.~\ref{fig:TimeSeries}a, the initial laser irradiation creates a vapor depression at time $t_0$ that deepens into the sample over the subsequent 350-$\mu$s. During this 3-frame progression, the vapor depression deepens by melting and displacing the molten metal toward the tail of the melt pool. The steep incline at the rear of the melt pool and surface tension initially pins this fluid in place as it accumulates. As time progresses however, still more molten metal is fed backward, beyond what surface tension can support causing the fluid flow to reverse toward the laser beam scanning direction, as demonstrated in Fig.~\ref{fig:TimeSeries}(b). Eventually this fluid flow reverses again as fluid is removed from the melt pool by balling and surface tension can accommodate greater fluid volumes. In some cases, the flow direction reversal continues periodically to create oscillatory chevron features of large amplitude.
As the fluid flows away from the laser beam, parallel to the laser beam scanning direction, we observe that the initially unstable vapor depression morphology stabilizes.  At $t_0 + 850 \mu s$ (Fig.~\ref{fig:TimeSeries}(b)), we observe the volume of the balled region to continue increasing as the laser beam advances. The advancing laser beam melts more metal and directs the fluid with high velocities toward the balling region. The fluid transport occurs through a fluid channel, as shown in Fig.~\ref{fig:TimeSeries}(b). By the end of the 3-frame progression in Fig.~\ref{fig:TimeSeries}(b), the fluid channel narrows, creating a ``necking'' region that separates the molten material into two separate regions. Following that progression, the subsequent image sequence in Fig.~\ref{fig:TimeSeries} (c) demonstrates how the solidification at the necking region prevents further growth of the balling region and repeats the process in the next region of material. 

\begin{figure}
    \centering
    \includegraphics[width=\linewidth]{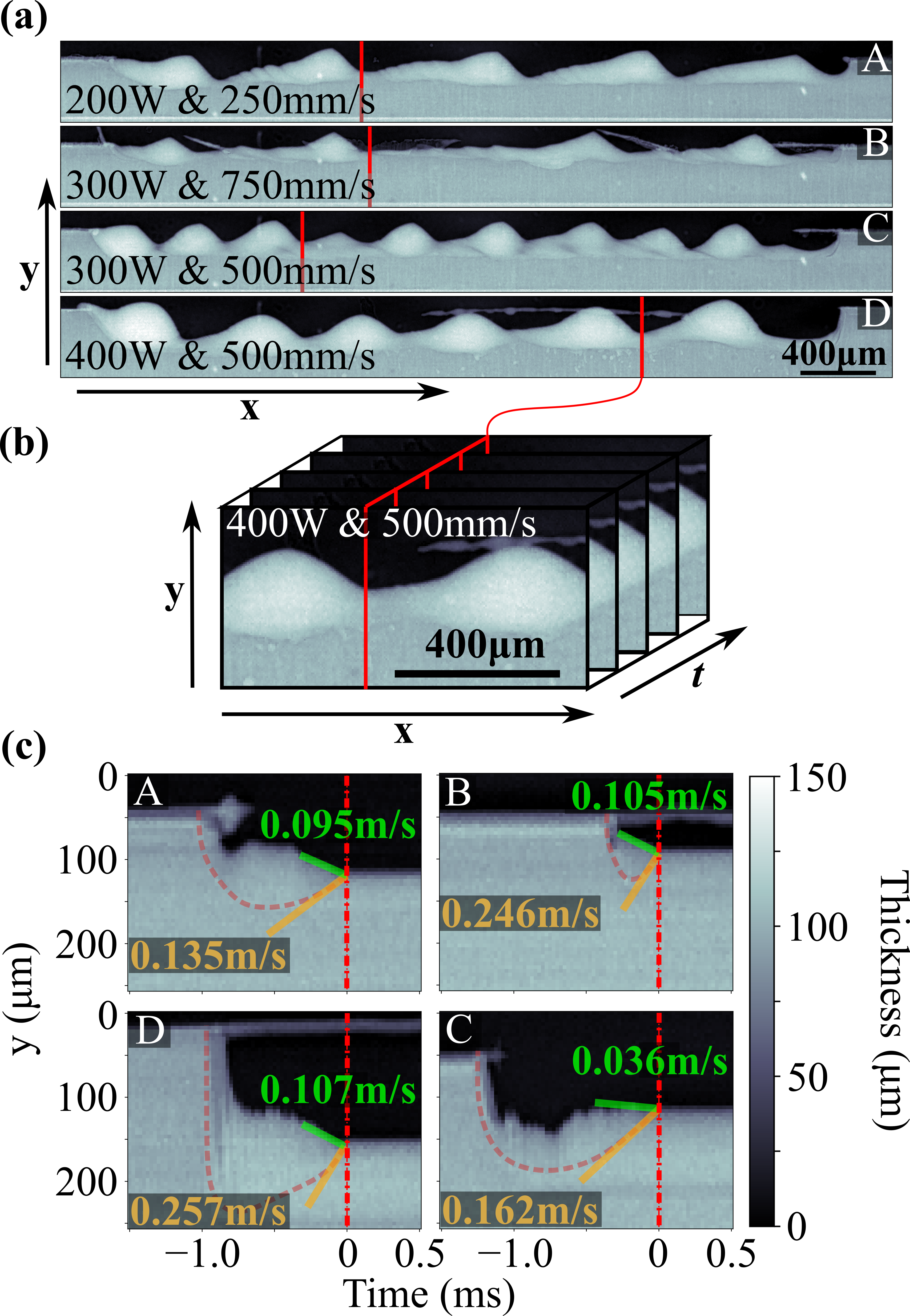}
    \caption{(a) Thickness plots after laser scan completion. (b) Datacube schematic of how waterfall plots are constructed and (c) the time progression of the sample cross-section at a the red lines in (a), the valley between balling events. The vertical red dashed line in (c) demarcates the time at which solidification pinching occurred, and the red curved dashed line shows the melt pool profile at that position over time.}
    \label{fig:waterfall}
\end{figure}

In Fig.~\ref{fig:TimeSeries}(c), the solidification front supports the formation of the next ball, while the solidified channel separates the previously molten region. As the previous hill still contains molten metal, it then undergoes relatively slow solidification due to its extra thermal mass, as highlighted with red dotted lines in the subsequent frames. The process of fluid channel narrowing and solidifying is clearly shown in the second and third frames of Fig.~\ref{fig:TimeSeries}(c). As the solidification front progresses toward the fluid channel, the top surface of the channel is driven downward by fluid instabilities. Once the fluid channel solidifies, fluid immediately accumulates at a new location. Initially, this accumulation is supported by the recently solidified fluid channel. The process defined in Fig.~\ref{fig:TimeSeries} repeats cyclically along the laser track, leading to the periodic structure characteristic of the balling defect.

The behavior of the fluid channel as it solidifies is revealed through time-domain waterfall plots in Fig.~\ref{fig:waterfall}. The waterfall plots of the channels marked in red in Fig.~\ref{fig:waterfall}(a) are presented in Fig.~\ref{fig:waterfall}(c), showing how that column of pixels varies over time. The top surface has a relatively complex behavior depending on the fluid oscillations of the channel but converges downward towards the end of solidification as seen in Fig.~\ref{fig:waterfall}(c). In contrast, the melt pool boundary at that location first drops as the laser passes and then rises upwards during solidification. The tangent lines and speeds of these boundaries at the end of solidification are also given in Fig.~\ref{fig:waterfall}(c), which reveal how solidification and the motion of the fluid surface are of comparable magnitudes and thus compete.

In Fig.~\ref{fig:waterfall}(c), we observe the vertical downward component of the free surface and the vertical upward component of the melt pool boundary. The vertical solidification velocity shows dependence on both laser power and scan velocity, increasing with higher values of each, as known from welding literature \cite{Rappaz1989}. For example, the solidification velocity (from the tangent slopes of the solid-liquid interface in Fig. \ref{fig:waterfall}(c)) increases from 0.162 m/s to 0.246 m/s as the scan velocity increases from 500 mm/s to 750 mm/s at a constant power of $P = 300$ W. Similarly, the solidification vertical velocity rises from 0.162 m/s to 0.257 m/s as laser power increases from 300 W to 400 W at a constant scan velocity of $v = 500$ mm/s. In contrast, the free surface downward velocity does not exhibit a clear correlation with either $P$ or $v$. The significant finding from these velocity measurements is that the magnitudes of both the melt pool solidification and free surface velocities are substantial. As a result, neither can be neglected when modeling balling behavior in AM. Therefore, in the following sections, we develop a formalism to account for the competition between solidification and the free surface dynamics driven by fluid instabilities.

\section{\label{sec:instability-theory}Review of Instability Theory}

\begin{figure*}
    \centering
    \includegraphics[width=0.8\linewidth]{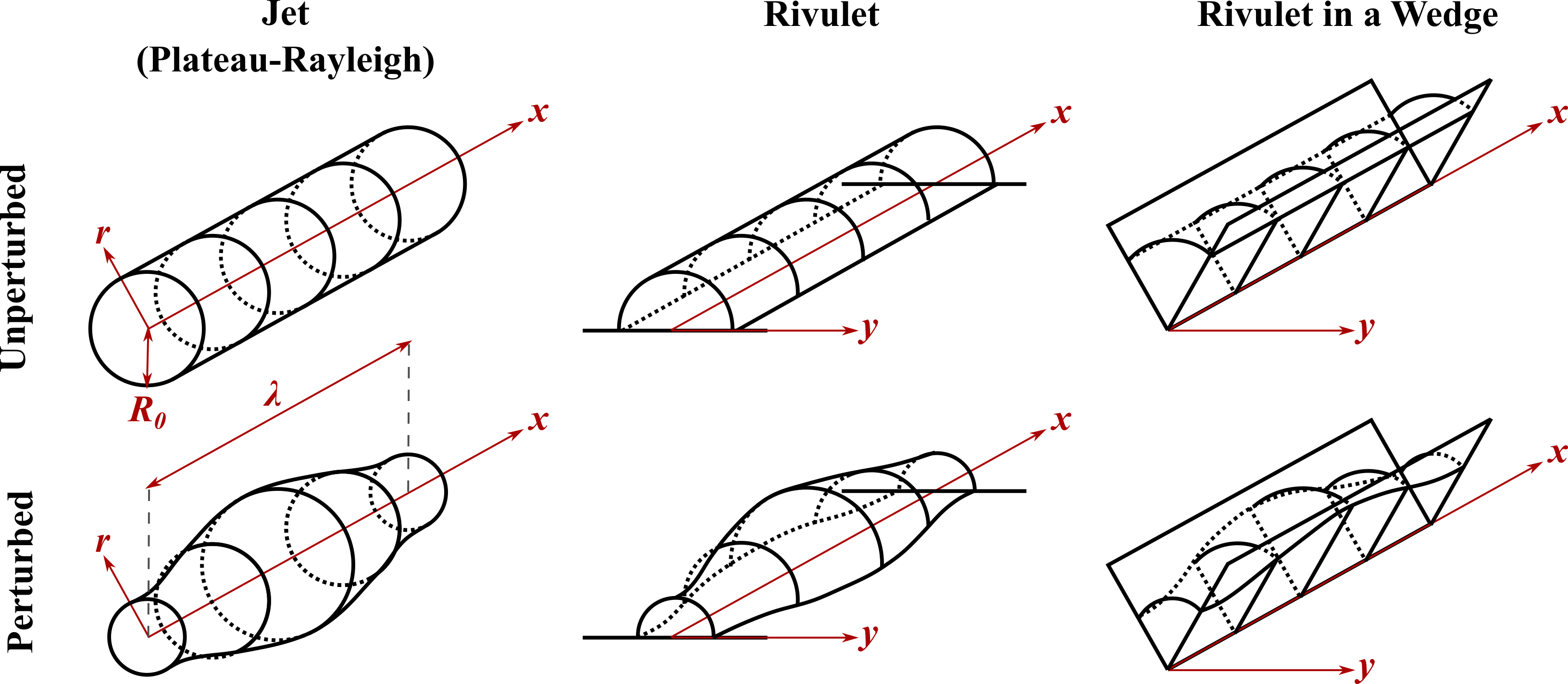}
    \caption{Progression from the cylindrical fluid jet in the absence of other media assumed by Plateau-Rayleigh theory, to a rivulet flowing on a substrate, to the rivulet supported by a wedge, more closely resembling a melt pool situation.}
    \label{fig:morphology-progression}
\end{figure*}

The Plateau-Rayleigh instability was first experimentally observed in 1873 and later theoretically explored in the fluid mechanics community \cite{McCuan1997}. The Euler equation is used to describe the fluid flow, which assumes an adiabatic and inviscid fluid. Further assumptions of incompressibility and irrotational fluid flow are imposed to ensure the fluid is not behaving turbulently and solve the Euler equation analytically \cite{Drazin2004}. Since these equations are linear and separable, the fluid instability community evaluates the stability of the fluid surface for only one spatial frequency of the perturbation at a time, analogous to a Fourier transform.

The cylinder of fluid considered in Plateau-Rayleigh is termed a ``jet" with initial radius $R_0$ that is sinusoidally perturbed by a perturbation
\begin{equation}
    \eta(x, t) = \eta_0 e^{ikx-i\Omega t},
    \label{eq:perturbation}
\end{equation}
where $\eta_0$ is the initial perturbation amplitude, $k=2\pi/\lambda$ is the wavenumber describing the spatial frequency of this standing wave under consideration, with wavelength $\lambda$ and temporal frequency $\Omega$, containing both real (oscillatory) and imaginary (growth) components.

It has previously be shown in literature \cite{Drazin2004, Batchelor1973} that with the assumptions defined above, the dispersion relationship between $\Omega$ and $k$ which describes the perturbation is given by
\begin{subequations}
\begin{align}
    \Omega^2 &= \frac{k\sigma}{\rho R_0^2}\cdot\frac{I_1(kR_0)}{I_0(kR_0)}(k^2R_0^2-1) \\
    \Omega^2 &\sim k^2 R_0^2 -1
    \label{eq:PR-dispersion}
\end{align}
\end{subequations}
for a fluid of density $\rho$, surface tension $\sigma$ and initial jet radius $R_0$. In this solution, $I_0$ and $I_1$ are the modified Bessel functions resulting from the solution for the internal pressure gradient.

The formula from Eq.~\ref{eq:perturbation} describes how the amplitude of the perturbation will grow exponentially when $\Omega$ has an imaginary component ($\Omega^2<0$ or $\Im\Omega\neq0$). In this work, we define this growth rate as $\omega=\Im\Omega$ to reconcile the notational differences between Plateau-Rayleigh and Rivulet theories in literature. Thus, $\omega>0$ defines the instability condition, 
\begin{equation}
    kR_0=\frac{2\pi}{\lambda}R_0=\pi\frac{2R_0}{\lambda}<1.
    \label{eq:PR_instability_condition}
\end{equation}

The AM community defines the melt pool width as a constant $W=2R_0$ and makes the mathematically convenient (though not necessarily valid) assumption that the wavelength is equal to the melt pool length, $\lambda=L$, so the resulting $L/W>\pi$ criterion commonly reported may be derived from Eq.~\ref{eq:PR_instability_condition}.

In contrast to the Plateau-Rayleigh instability --- developed to address the behavior of a cylindrical jet of fluid in isolation from all surfaces --- rivulet instability theory considers a fluid rivulet (instead of a jet) in contact with a surface, as shown schematically in Fig.~\ref{fig:morphology-progression}. The theoretical community exploring Rivulet instability is relatively small and typically explores flow on flat surfaces, varying assumptions with and without gravity, van der Waals forces, viscosity, and Laplace pressure, and using a wide variety of techniques, including lubrication theory and contact line perturbation \cite{Diez2009}. Since the Rivulet instability theory is more developed and more physically relevant to the AM geometry, we focus our work on their models rather than making substantial modifications to Plateau-Rayleigh treatment.

The form of Rivulet theory we consider in this work was developed by Yang and Homsey \cite{Yang2007, Yang2006} to describe a more general case of a rivulet flowing in a wedge. Their work assumes laminar fluid flow along the wedge and negligible gravitational effects. They further employ lubrication theory, which assumes that at least one of the dimensions is much smaller than the others; this enables a Taylor expansion of the Navier-Stokes equations along the small dimensions, offering an analytical solution \cite{Spurk2020}. The fundamental assumptions of this model are the quasiparallel, quasistatic flows and relatively long wavelengths for the resulting instabilities \cite{Yang2007}. Unlike other theories, Yang and Homsey use both axial and longitudinal curvatures to derive the capillary pressure inside the rivulet. They further assume Poiseuille flow --- laminar flow of an incompressible, Newtonian fluid \cite{Spurk2020} --- to relate this pressure gradient to volume flow rate.

The model of Yang and Homsey builds upon insights from the work of \cite{Ransohoff1988}, which describe fluids flowing in a V-shaped corner with varying degrees of roundedness. Together, these models describe a fluid which is supported from below and on its sides by the ``wedge", which in the case of AM by LPBF is the unmelted substrate.

\begin{figure}
    \centering
    \includegraphics[width=0.8\linewidth]{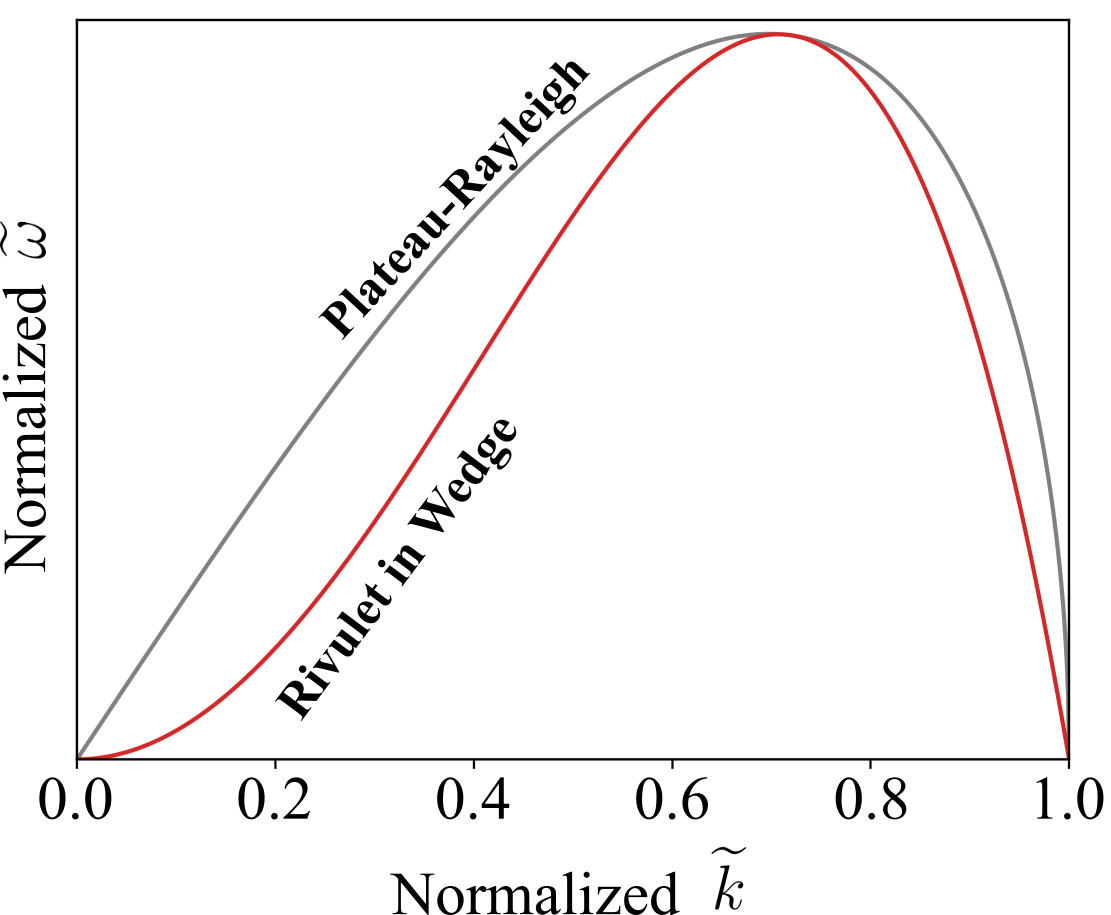}
    \caption{Comparison of the normalized dispersion relationships between Plateau-Rayleigh, Eq. \ref{eq:PR-dispersion}, and Rivulet in a wedge, Eq. \ref{eq:rivulet-omega}.}
    \label{fig:dispersion-comparison}
\end{figure}

Lang and Homsey's theory solves the fluid instability dispersion relation (analogous to Eq. \ref{eq:PR-dispersion}), describing the growth rate as
\begin{equation}
    \widetilde{\omega} = \widetilde{k}^2 - \widetilde{k}^4
    \label{eq:rivulet-omega}
\end{equation}
where $\widetilde{k}=kz_0$ is the dimensionless wavenumber, $\widetilde{\omega}=\omega\widetilde{t}_0$ is the dimensionless growth rate, $\widetilde{t}_0=a_0t_0$ is the normalized time constant, $z_0=a_0C$ is the geometry-dependent spatial constant, $a_0$ is the wedge radius, and $C$ is a factor depending on rivulet geometry. We note that the notation for the Lang and Homsey formalism includes the scaling factor of $z_0$ and $t_0$ to convert the observables predicted by e.g. the Plateau-Rayleigh theory into dimensionless numbers. Fig. \ref{fig:dispersion-comparison} compares the normalized dispersion relationships between the Plateau-Rayleigh theory with $\widetilde{k}=kR_0$ and the Rivulet theory of Eq. \ref{eq:rivulet-omega}. In normalized terms, the Rivulet instability has slower growth rates relative to the Plateau-Rayleigh instability.

\begin{figure}
    \centering
    \includegraphics[width=0.8\linewidth]{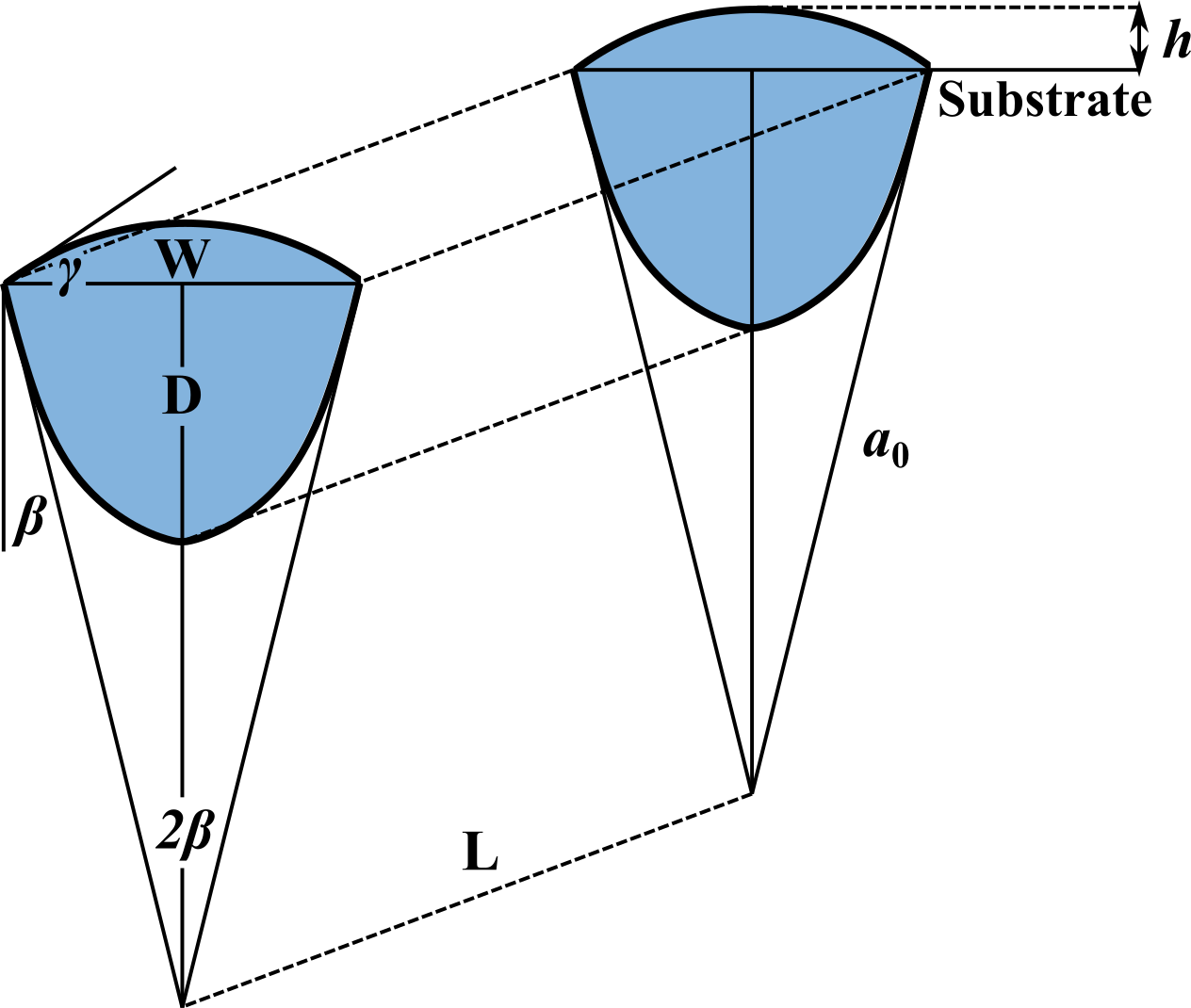}
    \caption{Schematic diagram of rivulet as an approximation of the AM melt pool. The melt pool has width $W$ and depth $D$ approximated as a parabolic shape, length $L$, a fluid-filling angle $\gamma$ above the substrate surface, and melt pool steepness at the surface $\beta$ which describes the angular width of the wedge containing the melt pool.}
    \label{fig:rivulet-definition}
\end{figure}

In this work, we modify the Yang and Homsey \cite{Yang2007} geometrical convention from being based on the wetting angle, $\theta$, to instead define the geometry based on the angle above the substrate surface $\gamma=\theta+\beta-\pi/2$ as shown in Fig.~\ref{fig:rivulet-definition}. We change this convention because $\gamma$ is more readily measurable in experiments and wetting angle hysteresis allows for multiple values of $\theta$, introducing ambiguity. This results in spatial and temporal scaling equations of the form,
\begin{subequations}
\begin{align}
     C&=\sqrt{-\frac{B}{K}}\\
     t_0&=\frac{\mu}{\sigma}\cdot\frac{EB}{K^2Q}
\end{align}
\end{subequations}
where $K<0$ for all reasonable values of $\gamma$ and the following geometric functions describe the rivulet
\begin{subequations}
\begin{align}
    B&=\frac{-\gamma\cos(\gamma+\pi/2-\beta)-\sin\beta\cos(\gamma+\pi/2)}{2\cos^2(\gamma+\pi/2)}\nonumber\\
    &+\frac{\sin(\gamma+\pi/2-\beta)}{2}\nonumber\\
    &=\frac{\gamma\sin(\gamma-\beta)+\sin\beta\sin\gamma}{2\sin^2\gamma}+\frac{\cos(\gamma-\beta)}{2}\\
    K&=\frac{\cos(\gamma+\pi/2)}{\sin\beta}=-\frac{\sin\gamma}{\sin\beta}\\
E&=2\sin\beta\cos\beta+2\tan(\gamma+\pi/2)\sin^2\beta+2\gamma\frac{\sin^2\beta}{\cos^2(\gamma+\pi/2)}\nonumber\\
    &=\sin2\beta-2\frac{\sin^2\beta}{\tan\gamma}+2\gamma\frac{\sin^2\beta}{\sin^2\gamma}
    \label{eq:rivulet-functionals}
\end{align}
\end{subequations}
where $a_0$ is the radius and $\beta$ is the half-angle of the wedge describing the melt pool, related to the steepness of the melt-pool wall at the surface, $\mu$ is the fluid viscosity and $\sigma$ is the fluid surface tension, and $Q$ describes the dimensionless volumetric flow of the melt pool geometry. Note that $\omega$ has units of $s^{-1}$ and $k=2\pi/\lambda$ has units of $m^{-1}$. From Yang and Homsey's interpretation, $K$ represents the contribution of the curvature of the meniscus (positive when the wedge is underfilled), $B$ is related to the variation or curvature along the wedge, and $E$ is related to the cross-sectional area of the rivulet \cite{Yang2007}. Note that these authors consider a constant-temperature case in which the fluid is not actively solidifying, whereas laser welding and AM fluid temperatures vary over a wide range during the solidification process. This model only accounts for balling effects that are zero-order in the temperature scaling of $\mu$ and $\sigma$ and as such we will consider these values taken at the material's melting temperature.

The Yang and Homsey version of the Rivulet model defines the condition for $\omega>0$ as the fluid instability criterion for 
\begin{equation}
    \widetilde{k}<1,
    \label{eq:rivulet-stability-criterion}
\end{equation}
which occurs when
\begin{equation}
    kz_0 = ka_0C = \frac{2\pi}{\lambda} a_0 C < 1  \hspace{0.5cm} \text{or} \hspace{0.5cm} 2\pi C < \lambda/a_0.
    \label{eq:stability-criterion}
\end{equation}
The model further defines a maximum growth rate $\omega$ occurring at
\begin{subequations}
\begin{align}
    \widetilde{k} &= \frac{1}{\sqrt{2}}\\
    \lambda_{max}/a_0 &= \sqrt{2}\pi C\\
    \omega_{max} &= \frac{1}{4a_0t_0}
\end{align}
\end{subequations}
In this section, we've presented established Plateau-Rayleigh theory from literature to show how the AM community obtains its threshold criteria for balling using stability analysis of their respective dispersion relationships. We went on to present Yang and Homsey's Rivulet instability theory for the special case of a rivulet in a wedge, which more closely resembles realist laser melting and AM conditions.

\section{\label{sec:model-development}Model Development:\protect\\ Competition between Fluid Instability and Solidification}

In this section, we describe our model developments. First, the adaptation of the rivulet-in-a-wedge instability theory introduced in Section \ref{sec:instability-theory} to the AM condition. We then introduce a formalism to account for the competition of solidification and fluid instability through their relative timescales, enabling us to derive a gradual transition from normal printing to balling, instead of the binary transition predicted previously.

\subsection{\label{sec:adapting-rivulet-to-AM}Adapting Rivulet Instability to AM}

We begin by extending the geometry of the Yang and Homsey Rivulet model to the AM system using the definitions shown schematically in Fig.~\ref{fig:rivulet-definition}. In this formalism, $a_0$ is the radius of the triangular wedge that circumscribes the melt pool and can be related to the melt pool width $W$ as $a_0=W/2\sin\beta$, and $D$ is the melt pool depth, which does not extend to the bottom of the wedge. The height of the melt pool above the substrate $h$ is described by the angle $\gamma$, and in the absence of gravity, $h=(W/2)\tan(\gamma/2)$, valid for small melt pools common in LPBF. For generality, we let $\lambda\neq L$. In a later section and the Supplemental text, we describe two formalisms by which this inequality is a natural result.

With these definitions substituted into Eq. \ref{eq:rivulet-omega}-\ref{eq:rivulet-functionals}, we get the result that 
\begin{subequations}
\begin{align}
    \widetilde{k}&=kz_0=\frac{2\pi}{\lambda}\cdot a_0C=\frac{2\pi}{\lambda}\cdot\frac{W}{2\sin\beta}C=\frac{\pi C}{(\lambda/W)\cdot\sin\beta}\\
    \widetilde{t}_0&=a_0t_0=\frac{W}{2\sin\beta}t_0
\end{align}
\end{subequations}
which are the dimensionless wavenumber and normalized time constant with respect to the melt pool geometry.

If we further assume that the cross section of the melt pool is described by a parabola (a common assumption in literature \cite{Seede2020}) of width $W$ and depth $D$ given by the function $y=D(1-4z^2/W^2)$, the slope of the parabola when it intersects the substrate surface defines the wedge encompassing the melt pool from Rivulet theory, i.e.,
\begin{equation}
    \beta = \tan^{-1}\bigg(\frac{W}{4D}\bigg).
    \label{eq:beta}
\end{equation}

Our treatment of the melt pool geometry does not explicitly account for a powder layer. Implicitly, powder incorporation is the source of fluid entering the melt pool and results in $h>0$ and $\gamma>0$ in standard printing conditions. The more powder is incorporated into the melt track, the larger $h$ and $\gamma$ will be. In the Supplemental text, we provide an expression to estimate $h$ and $\gamma$ from the powder layer parameters. Outside of how powder incorporation influences $h$ and $\gamma$ of the melt pool geometry, we assume powder is non-interacting with the Rivulet instability and does not contribute any support like the wedge. A result of our powder assumptions is that the regime of balling under consideration has sufficient laser energy density to penetrate through the powder layer into the substrate. If penetration were not the case, the melt pool would more closely resemble the original Plateau-Rayleigh formalism as a ``cylinder" suspended in a powder layer rather than a rivulet on a substrate surface. Consideration of the partial wetting of powder is complex beyond the scope of this work, but qualitatively the presence of powder could be taken to influence the initial perturbation amplitude $\eta_0$ by disturbing the fluid surface. 

\begin{figure}
    \centering
    \includegraphics[width=0.6\linewidth]{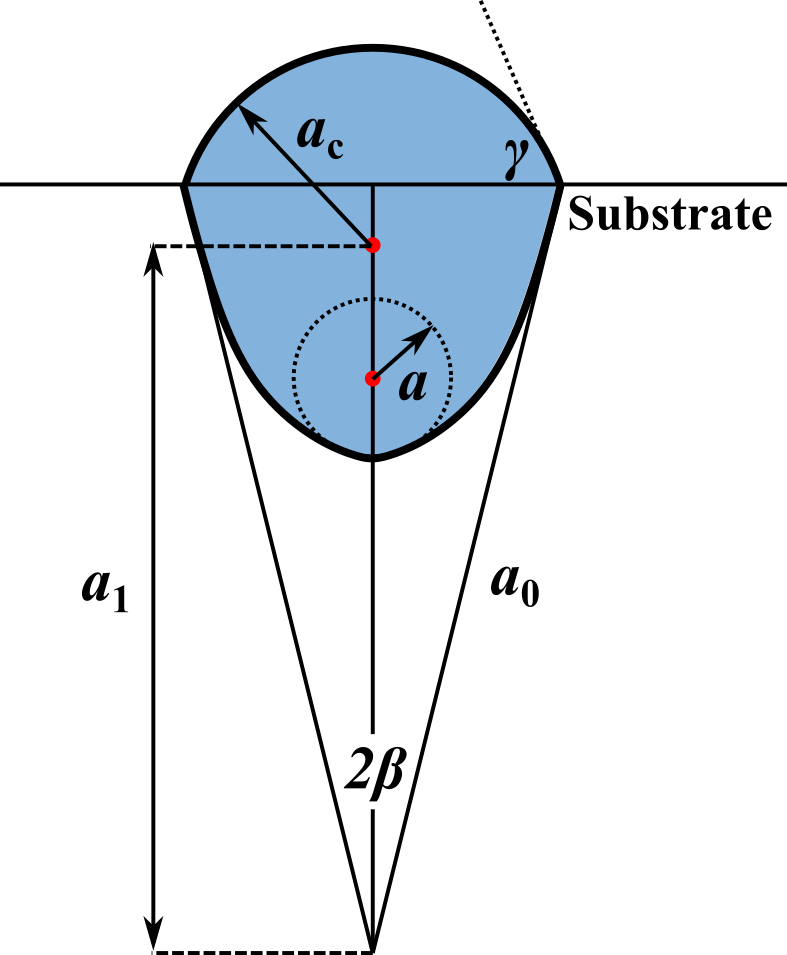}
    \caption{Definitions for the roundedness parameter $\mathfrak{r}$.}
    \label{fig:roundedness-cartoon}
\end{figure}

The final consideration necessary to extend the Yang and Homsey Rivulet instability model to AM conditions is to account for the roundedness of the melt pool, depicted in Fig. \ref{fig:roundedness-cartoon}. Yang and Homsey's theory uses the dimensionless volumetric flow, $Q$, which is a function of order $10^{-4}$. This parameter was earlier proscribed by Ransohoff and Radke and is related to the roundedness of the wedge \cite{Ransohoff1988}. Specifically, they define
\begin{equation}
    Q=Q_0(1-\mathfrak{r})^3
    \label{eq:Q_def}
\end{equation}
in the limit of large roundedness, $\mathfrak{r}\to1$, with a prefactor of the order $Q_0\sim10^{-4}$. $Q_0$ is reported to have some dependence on the wedge geometry and contact angle, but the variation is of order unity and $Q_0$ is independent of fluid medium by construction so this value should be applicable to AM conditions as well. The degree of roundedness $\mathfrak{r}$ is defined as,
\begin{equation}
    \mathfrak{r}=\frac{a_1-a}{a_1-a_c}\in[0,1].
    \label{eq:roundedness}
\end{equation}
$a_1$ is the distance from the center of the circle defining the liquid-gas interface to the corner of the wedge. $a$ is the distance from the center of the circle defining the liquid-gas interface to the edge of the rounded part of the corner, which we approximate as the curvature of the wedge corner. Finally, the term $a_c$ is the radius of curvature of the circle defining the liquid-gas interface. These definitions are shown schematically in Figure \ref{fig:roundedness-cartoon}.

\begin{figure*}
    \centering
    \includegraphics[width=0.8\linewidth]{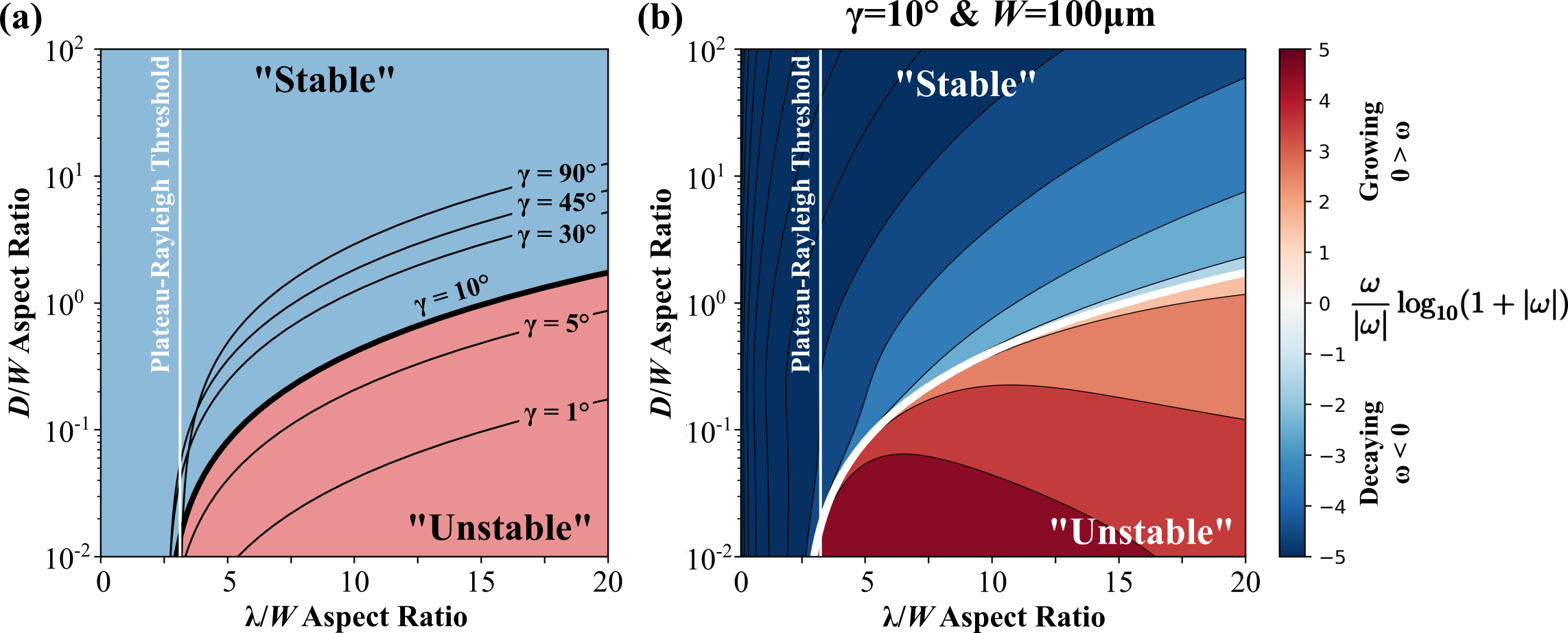}
    \caption{(a) Stability diagram for melt pool morphology, compared to the Plateau-Rayleigh threshold in white. Blue is stable and red is unstable for $\gamma=10^\circ$. Threshold isolines are shown for various $\gamma$ values. (b) Fluid instability growth rate $\omega$ ($s^{-1}$) as a result of melt pool morphology, calculated using rivulet theory. The white curve is the $\omega=0$ line. Results are plotted as a signed psuedo-logarithm to show positive and negative growth over many orders of magnitude.}
    \label{fig:binary_threshold}
\end{figure*}

In our laser welding or AM case, $a_1=\frac{W}{2}(\cot\beta-\cot\gamma)$ is the distance between the center of the circle defined by the melt pool's top surface and the location of the wedge corner that is defined by the edges of the melt pool's slope. The radius of curvature of a parabola with width $W$ and depth $D$ is $a=W^2/8D$, and the liquid-gas curvature is $a_c=-W/2\sin\gamma$ which is negative because the fluid concavity is the opposite of that considered in Ransohoff and Radke's work. We incorporate these AM attributes into the roundedness factor from Eq.~\ref{eq:roundedness} giving
\begin{align}
    \mathfrak{r}&=\frac{\frac{W}{2}(\cot\beta-\cot\gamma)-\frac{W^2}{8D}}{\frac{W}{2}(\cot\beta-\cot\gamma)+\frac{W}{2\sin\gamma}}\nonumber\\
    &=\frac{(\cot\beta-\cot\gamma)-\frac{W}{4D}}{(\cot\beta-\cot\gamma)+\frac{1}{\sin\gamma}}\nonumber\\
    &=\frac{\tan\gamma-\tan\beta-\tan^2\beta\tan\gamma}{\tan\gamma-\tan\beta+\frac{\tan\beta\tan\gamma}{\sin\gamma}}\\
    \label{eq:roundedness-explicit}
    &\approx 1-\frac{\beta}{\gamma}\nonumber\\
    1-\mathfrak{r}&\approx\frac{\beta}{\gamma},
\end{align}
where we have reported the small angle approximation for $\beta$ and $\gamma$. Using the small angle approximation for $\mathfrak{r}$ in Eq. \ref{eq:Q_def} gives
\begin{equation}
    Q\approx 10^{-4}\cdot\Bigg(\frac{\beta}{\gamma}\Bigg)^3,
    \label{eq:Q_approx}
\end{equation}
which is valid for deep melt pools with relatively thin build layers. In the remainder of this work, however, we use the exact definition of $\mathfrak{r}$ to allow us to explore the effects of shallow melt pool geometries and large $\gamma$ angles. In this work, we will also consider $Q_0=10^{-4}$ a fixed parameter instead of calculating it using fluid dynamics simulations or treating it as a fitting parameter. If future work wished to treat $Q_0$ as a fitting parameter, it would be geometry dependent but material agnostic.

The physical meaning of $Q$ is a dimensionless flow rate and $1/Q$ is a dimensionless flow resistance. $Q_0$ is the value of this $Q$ when the wedge comes to a sharp corner. When the wedge is shallower with respect to its depth, or $\beta$ increasing, the parabola defining the melt pool occupies more of the cross-sectional area of the wedge, so the flow rate relative to the sharp wedge increases and the flow resistance decreases. Conversely, increasing $\gamma$ or powder incorporation, decreases flow rate and increases flow resistance through the wedge since the fluid flows less through the wedge and more through the section of the melt pool above the substrate surface.

Now that we have defined $a_0$, $\beta$, and $Q$ in terms of the melt pool geometry in the laser melting case for instability theory with a rivulet in a wedge, we may consider the stability of a melt pool given its geometry.

Evaluating the Rivulet stability criterion in Eq. \ref{eq:rivulet-stability-criterion}, we obtain the stability diagram in Fig. \ref{fig:binary_threshold}(a) which is a function of the melt pool geometry in its wavelength-to-width aspect ratio as well as its depth-to-width aspect ratio. While $\lambda\neq L$, the melt pool length does place a bound on how long $\lambda$ can be. In contrast, the Plateau-Rayleigh stability threshold is a constant value independent of the depth-to-width aspect ratio of the melt pool. The stability criterion here is a binary threshold in the same way the AM community uses the Plateau-Rayleigh stability criterion with $\omega>0$ being unstable, and would be valid if the solidification rate was negligible. Fig. \ref{fig:binary_threshold}(a) shows how the melt pool becomes more unstable as the layer height increases ($\gamma$ increasing). Larger $\gamma$ is destabilizing because there is greater fluid surface area. Fig. \ref{fig:binary_threshold}(a) further reveals that a deeper melt pool (relative to the width) is more stable than a shallower one due to the stabilizing presence of the substrate support. The depth-to-width ratio's influence on melt-pool stability helps explain a commonly observed balling trend: as the laser scan velocity increases, the depth-to-width decreases, lowering the critical threshold for balling and decreasing the wavelength of the observed balling structure.

Note that our analysis is scale invariant, with the melt pool geometry normalized to its width, and since the $\omega=0$ lines define the Rivulet stability criterion in Eq. \ref{eq:rivulet-stability-criterion}, the timescale and growth rates do not factor into the expression. The scale of the system, $W$, does not affect where the critical threshold for unstable (positively growing) melt pool geometries lies, but does effect the timescale at which it happens. 

As soon as the melt pool scale $W$ and fluid overfilling $\gamma$ are fixed, the fluid instability growth rate $\omega$ can be calculated explicitly. Such a calculation is shown in Fig. \ref{fig:binary_threshold}(b) for a 100$\mu$m wide melt pool with $\gamma=10^\circ$, depicting a contour plot of the growth rate $\omega$. The $\gamma=10^\circ$ curve of Fig. \ref{fig:binary_threshold}(a) is equivalent to the white $\omega=0$ curve of Fig. \ref{fig:binary_threshold}(b), confirming that the binary threshold between positive and negative fluid instability growth rate is scale invariant. For wavelengths longer than a critical length, $\omega>0$ and the rivulet is unstable, but the magnitude of that instability growth rate is greater for smaller depth-to-width aspect ratio melt pools. Conversely, wavelengths shorter than the critical length are stable with $\omega<0$ and the magnitude of the instability decay is greater for shorter and deeper melt pools.

In this subsection, we have addressed how the Lang and Homsey Rivulet instability model can be recast using the AM geometry and conventions to obtain both binary stability diagrams and fluid instability growth rates as a function of the melt pool's 3D geometry.

\subsection{\label{sec:competition}Competition between Fluid Instability and Solidification}

In this subsection, we define a formalism to describe the competition between the fluid instabilities discussed in Section~\ref{sec:instability-theory} and the solidification front that defines the lower boundary of the melt pool. We do this by calculating the point in time at which the two surfaces meet, fragmenting the melt pool. At this point, fluid can no longer flow backwards in the melt pool and the geometry of the isolated melt pool fragment is largely fixed, forming the ``hills" that are observed experimentally in the balling process. Depending on the relative kinetics of the solidification and the fluid instability, the depth at which the two surfaces meet varies, controlling the amplitude of the balling. 

\begin{figure}
    \centering
    \includegraphics[width=\linewidth]{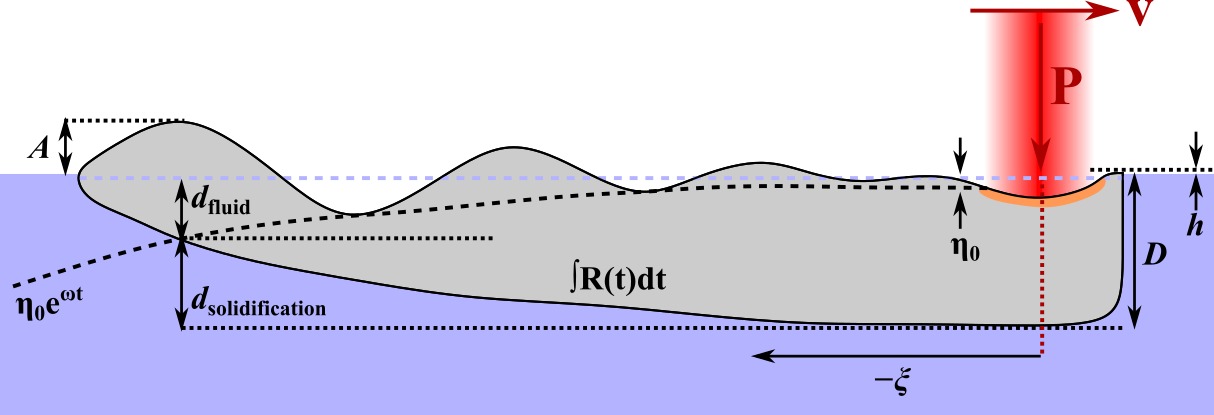}
    \caption{Evolution of the fluid instability surface, depicted as a sinusoid with growing amplitude, moving downwards toward the upward moving solidification front.}
    \label{fig:solidification-competition}
\end{figure}

In Fig. \ref{fig:solidification-competition}, the top surface of the fluid is subjected to a fluid instability and has a sinusoidal perturbation with an amplitude that increases over time. Because the laser is in motion, time $t$ can also be related to the distance in the melt pool behind the laser, $\xi=x-vt$. This amplitude can be thought of as an envelope function for the sinusoidal wave and the phase of that wave is constantly varying to ensure a continuous surface profile as the laser scans. As an equation, the top surface in 2D would be described by
\begin{equation}
    y(\xi) = h+A(\xi)e^{ik(x-\xi)}e^{i\phi},
    \label{eq:envelope-and-phase}
\end{equation}
where $h$ is the initial height above the substrate, $k=2\pi/\lambda$ is the wavenumber,  $A=\eta_0e^{\omega t}$ is the envelope function describing the fluid instability amplitude (recall Section \ref{sec:instability-theory}), $e^{ik(x-\xi)}$ describes how different locations along the melt track can be in a peak or valley of the sinusoid, and $e^{i\phi}$ is an arbitrary, constant phase shift. Since we assume balling by ``pinching" or melt pool fragmentation when the fluid surface and solidification front intersect, it is sufficient to consider only the valleys of the sinusoidal perturbation. Making this assumption sets $e^{ik(x-\xi)}e^{i\phi}=-1$ and reduces the competition between solidification and fluid instability to a 1D equation.

Pursuing this 1D equation, we describe the distance traversed by the fluid surface and solidification front as
\begin{subequations}
\begin{align}
    d_{\text{fluid surface}} &=\eta_0+ \int_0^\tau \eta_0\omega(R)e^{\omega(R)t} dt\\
    d_{\text{solidification front}} &= \int_0^\tau R(t)dt,
\end{align}
\end{subequations}
where $\tau$ is the completion time when the two surfaces intersect, $\omega$ is the fluid instability growth rate and $R$ is the solidification rate. This integral notation is intended to clarify that the fluid instability and solidification velocities are dependent on each other and the 3D melt pool geometry in the most general case. The condition for these surfaces to intersect and the fluid instability process to conclude is then given by,
\begin{align}
    D+h &= d_{\text{fluid surface}} + d_{\text{solidification front}}\nonumber \\
    &= \eta_0+\int_0^\tau \eta_0\omega(R)e^{\omega(R)t} + R(t) dt,
    \label{eq:D-h_balance}
\end{align}
where $D$ is the melt-pool depth and $h$ is the initial melt-pool height above the substrate surface. The solution of this equation may thus be highly nontrivial due to the underlying dependence of $\omega$ with $R$ and $t$. If we instead use the average fluid instability growth rate and average solidification rate over time at a fixed $x$ position, Eq. \ref{eq:D-h_balance} simplifies to 
\begin{equation}
    D+h=R\tau+\eta_0e^{\omega\tau}
    \label{eq:D-h_balance-simple}
\end{equation}
and we can solve the function in closed-form as
\begin{equation}
    \tau = \frac{D+h}{R}-\frac{1}{\omega}\text{ProductLog}\bigg(\eta_0\frac{\omega}{R}\exp\bigg[(D+h)\frac{\omega}{R}\bigg]\bigg),
    \label{eq:tau}
\end{equation}
where the $\text{ProductLog}$ function is also known as the Lambert W function, the inverse of $xe^x$ \cite{Mez2022}. In the remainder of this work, we will consider this $R$ as the \textit{average} vertical solidification rate and $\omega$ as the \textit{average} fluid instability growth rate. We choose the average rates to simplify the integrals in  Eq. \ref{eq:D-h_balance}, and the use of the vertical solidification rate stems from the 1D model simplification of Eq. \ref{eq:envelope-and-phase}. We also assume that this average $\omega$ is given by the initial melt pool geometry, independent of solidification. These assumptions are valid when solidification is slow relative to the fluid instability which corresponds to the balling conditions we aim to describe. The meaning of ``relatively slow" solidification depends on the value of $\omega$ resulting from the previous section and will become clear in following paragraphs.

\begin{figure}
    \centering
    \includegraphics[width=\linewidth]{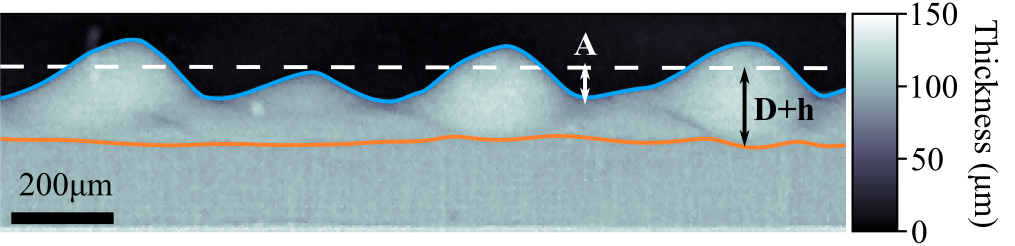}
    \caption{Radiograph in thickness units for 300W and 500mm/s with the absolute balling amplitude $A$ drawn over the top surface profile in blue, and the melt pool depth $D$ drawn in orange. Balling fraction $\Upsilon$ is the ratio of these terms.}
    \label{fig:balling-fraction}
\end{figure}

Recalling the balling amplitude referenced in Eq. \ref{eq:envelope-and-phase} and shown in Fig.~\ref{fig:balling-fraction},
\begin{equation}
    A = \eta_0+\int_0^\tau \eta_0\omega(R)e^{\omega(R)t}\approx\eta_0e^{\omega\tau}
\end{equation}
We further define a normalized balling amplitude, or the ``balling fraction" as
\begin{equation}
    \Upsilon = \frac{A}{D+h} \in [0, 1]
    \label{eq:upsilon_def}
\end{equation}
where $\Upsilon=0$ means that solidification dominates and there is no unevenness of the melt pool surface (neglecting chevroning), and $\Upsilon=1$ means that the fluid instability dominates and the entire depth of the melt pool contributes to $A$. With the assumptions made in Eq.~\ref{eq:tau}, $\Upsilon$ has the closed-form solution
\begin{align}
    \Upsilon = &\frac{\eta_0}{D+h}\exp\bigg[(D+h)\frac{\omega}{R}\bigg]\nonumber\\
    &\times\exp\bigg[-\text{ProductLog}\bigg(\eta_0\frac{\omega}{R}\exp\bigg[(D+h)\frac{\omega}{R}\bigg]\bigg)\bigg],
    \label{eq:upsilon}
\end{align}
which is a function of the initial perturbation amplitude of the fluid surface, $\eta_0$, for a given wavelength perturbation, the total initial melt pool depth, $D+h$, and the competition between fluid instability growth rates and solidification rates, $\omega/R$.

\begin{figure*}
    \centering
    \includegraphics[width=0.7\linewidth]{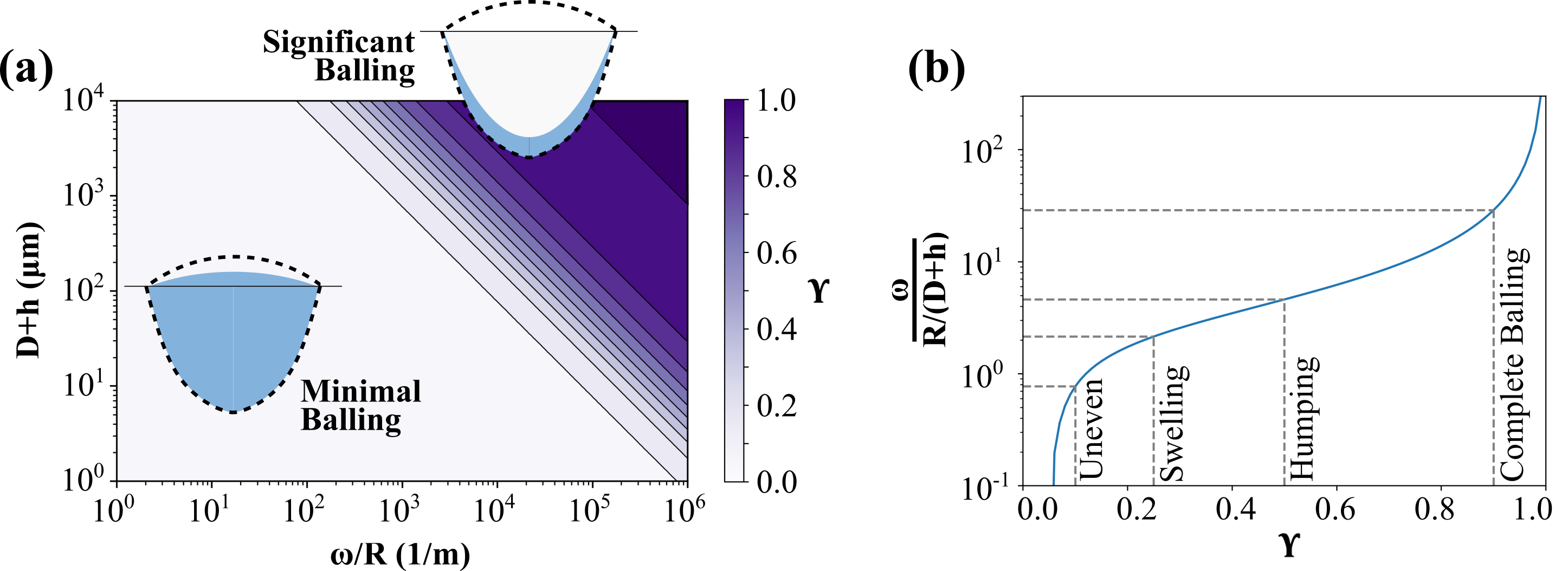}
    \caption{(a) Contour plot of the balling fraction $\Upsilon$ (assuming Eq. \ref{eq:upsilon}) in a space defined by the initial melt pool depth $D$ and the fluid instability growth rate relative to solidification rate, $\omega/R$.  Inlays compare the fluid solidified cross section at the pinch point. (b) Shows how the fluid instability growth rate relative to solidification competes with the balling fraction in a non-linear fashion. An initial $\eta_0/(D+h)=5\%$ is assumed for both calculations.}
    \label{fig:upsilon-D-vs-wR}
\end{figure*}

The behavior of Eq. \ref{eq:upsilon} is plotted in Fig. \ref{fig:upsilon-D-vs-wR} given some total melt pool depth and the fluid instability growth rate normalized to the solidification rate. Qualitatively, this means that deeper melt pools will experience more balling at a given solidification rate, because the melt pool will remain fluid for longer and give the fluid instability more time to grow. Note the apparent contradiction of this finding compared to the previous discussion where deeper melt pools stabilize the fluid. While true that deep melt pools have more stable $\omega$, \textit{for a given $\omega$ and $R$} deeper melt pools take longer to solidify favoring balling. Only by using both models in conjunction can an estimate be made of which effect dominates the balling phenomena. Similarly, at a given melt pool thickness, the larger the fluid instability growth rate is relative to the solidification velocity, the more balling will occur. This is depicted by the cartoon inlays in Fig. \ref{fig:upsilon-D-vs-wR}(a).

The function for $\Upsilon$ can be inverted to solve for either (a) the critical perturbation magnitude or (b) the critical fluid instability growth rate necessary to result in a particular balling fraction. These expressions are given as
\begin{subequations}
\begin{equation}
    \frac{\eta_0}{D+h}=\Upsilon_c\exp\Bigg[-(1-\Upsilon_c)(D+h)\frac{\omega}{R}\Bigg]
    \label{eq:upsilon-inverse-eta}
\end{equation}
\begin{equation}
    \frac{\omega}{R}=\frac{1/(D+h)}{1-\Upsilon_c}\ln\Bigg[\Upsilon_c\Bigg(\frac{D+h}{\eta_0}\Bigg)\Bigg],
    \label{eq:upsilon-inverse-omega}
\end{equation}
\end{subequations}
respectively. The intuition of Eq. \ref{eq:upsilon-inverse-eta} and Eq. \ref{eq:upsilon-inverse-omega} is similar to that of Eq. \ref{eq:upsilon}: faster growing fluid instabilities need smaller initial perturbations to obtain the same balling fraction. Eq. \ref{eq:upsilon-inverse-omega} is plotted in Fig. \ref{fig:upsilon-D-vs-wR}(b), to show the non-linear relationship between $\Upsilon$ and $\omega/R$.

For convenience and to add quantitative definitions for the nomenclature often used in the field, we define approximate thresholds in Table \ref{tab:balling_thresholds} that can be imposed by our model. These values are overlaid in Fig. \ref{fig:upsilon-D-vs-wR}(b). These values define the quantitative transition from initial melt pool unevenness to swelling to humping to complete balling, describing the gradual progression from stable melt pools to balling that are observed in other studies. We note that these transitions result from solidification timescales becoming slow with respect to the rate of the fluid instability.

\begin{table}
\caption{\label{tab:balling_thresholds}Arbitrary definitions for moderate and severe balling, in terms of their balling fraction, $\Upsilon$. Note that the last column is estimated assuming an initial $\eta_0/(D+h)=5\%$ perturbation.}
\begin{ruledtabular}
\begin{tabular}{rcc}
Condition & $\Upsilon_c$ & $\frac{\omega}{R/(D+h)}$\\
\hline
Unevenness & 0.10 & 0.77\\
Swelling & 0.25 & 2.15\\
Humping & 0.50 & 4.61\\
Complete Balling & 0.90 & 28.9\\
\end{tabular}
\end{ruledtabular}
\end{table}

We note that while more advanced models for $R$ may be implemented, which would allow Eq.~\ref{eq:tau} and \ref{eq:upsilon_def} to be evaluated numerically and more exactly, these values can also be measured from synchrotron radiography experiments or estimated as 
\begin{equation}
    R\approx\frac{D+h}{L}v \text{ or } \frac{R}{D+h} = \frac{v}{L}
    \label{eq:R_avg}
\end{equation}
where $R$ is the average vertical growth rate, $L$ is the melt pool length, and $v$ is the laser scan velocity.

\section{\label{sec:theory-to-experiments}Connecting Theory to Experiments}

To test our model's predictivity, we compare our findings to the Ta experiments introduced in Section \ref{sec:experimental-results}. We thus consider the material constants relevant to Ta, using the surface tension of $\sigma=2.1$ N/m and viscosity $\mu=8.5\times10^{-3}$ Pa$\cdot$s at the melting point \cite{Starodubtsev2021}.

\begin{figure}
    \centering
    \includegraphics[width=0.85\linewidth]{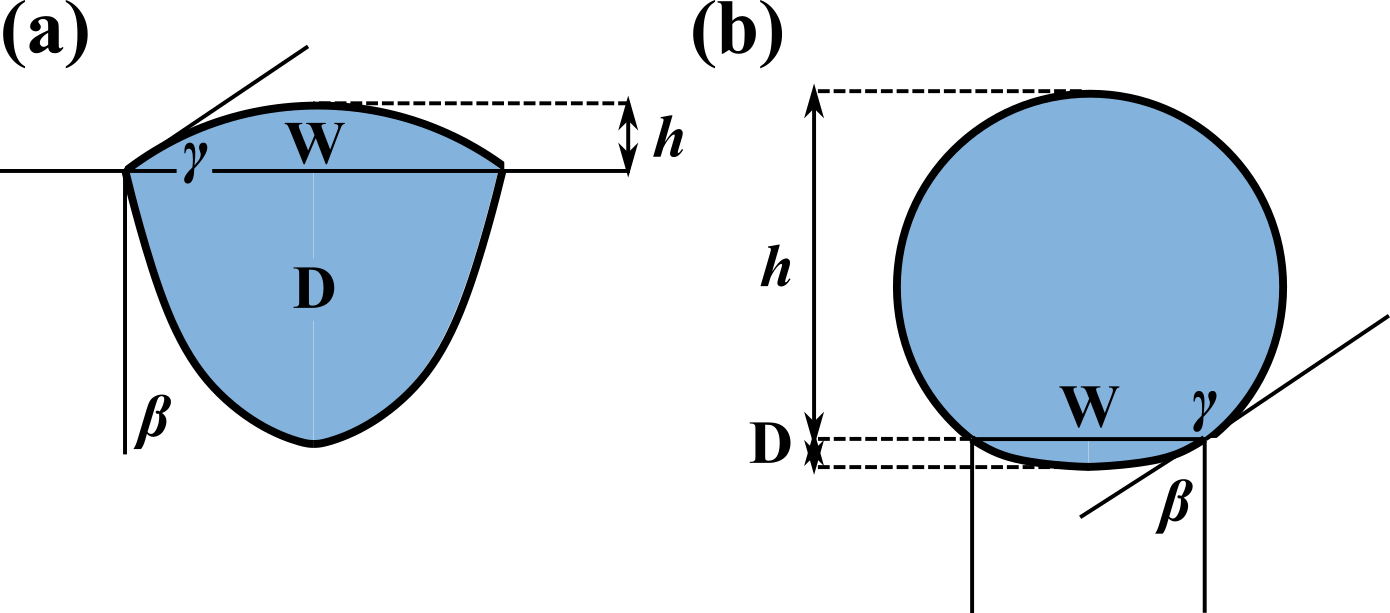}
    \caption{Schematic comparison of rivulet parameters in (a) normal melt pool supported in substrate vs (b) the case of a melt pool on a thin wall.}
    \label{fig:thin-wall-vs-normal}
\end{figure}

Since our radiography experiments required thin samples which melted through their entire thickness, there are no side walls supporting the melt pool; i.e., the melt pool is not in a wedge as we have depicted until now in Fig. \ref{fig:thin-wall-vs-normal}(a). Instead, as shown in Fig. \ref{fig:thin-wall-vs-normal}(b), the melt pool depth is shallow $D \ll h$ and $\beta\to\pi/2$. As can be predicted from Fig.~\ref{fig:binary_threshold}, the conditions from our experiments favor large and unstable growth rates for fluid instabilities, consistent with our imaging results.

However, the vanishing $D/W$ is not an issue when using the exact expressions for $\mathfrak{r}$ and $Q$, and the Rivulet instability model we have described is general enough to capture the odd case of our experimental conditions and evaluate the balling fraction without side-wall support. Our value for $W$ is set by the thin sample thickness, and from the thickness conversion of our radiography datasets, we can calculate the mean and standard deviation of the melt pool length $L$, thickness $D+h\approx h$, fluid angle $\gamma$ related to contact angle, balling periodicity $\lambda$, balling amplitude $A$, balling fraction $\Upsilon$, and various ratios thereof which are tabulated in Table \ref{tab:all-measured-parameters} for each case. Using waterfall plots, the solidification velocity at the end of solidification $R_t$, average solidification velocity $R$, and fluid surface velocity at the end of solidification $R_f$ are calculated and reported in Table \ref{tab:all-measured-parameters}. Note, the vertical solidification velocities are spatially non-uniform and much slower than the laser scan velocity due to the large angular differences between the laser scan direction and the melt pool surface normal. From our combined Rivulet-solidification model, we can also calculate the threshold wavelengths for which $\omega=0$ and fit $D/W$ and $\omega$ to the measured $\Upsilon$. Because $D/W$ is vanishing, it cannot be determined from the radiography and post-mortem analysis would be needed to estimate the value, which is beyond scope for this work. Since $D/W$ is the only unknown, it can be calculated using the measured $\Upsilon$, but we cannot independently estimate $\Upsilon$.

\begin{table}
\caption{\label{tab:all-measured-parameters}All measured parameters. Values are reported as the mean plus or minus the standard deviation.}
\begin{ruledtabular}
\begin{tabular}{l|cccc}
    Case & A & B & C & D \\\hline
    $P$ (W) & 200 & 300 & 300 & 400 \\
    $v$ (m/s)& 0.25& 0.75& 0.50& 0.50\\\hline
    $W$ ($\mu$m) & 100 & 100 & 100 & 100 \\
    $L$ ($\mu$m) & 355$\pm$111 & 351$\pm$147 & 508$\pm$200 & 539$\pm$217 \\
    $h$ ($\mu$m) & 72$\pm$14& 57$\pm$16& 52$\pm$14& 82$\pm$20\\
    $\gamma$ ($^\circ$) & 103$\pm$6.6& 97$\pm$5.9& 99$\pm$5.3& 105$\pm$7.9\\\hline
    $R_t$ (m/s) & 0.094$\pm$0.024 & 0.177$\pm$0.011 & 0.122$\pm$0.014 & 0.122$\pm$0.023 \\
    $R$ (m/s)& 0.080$\pm$0.025 & 0.190$\pm$0.080 & 0.123$\pm$0.048 & 0.147$\pm$0.059\\
    $R_f$ (m/s) & 0.062$\pm$0.038 & 0.099$\pm$0.051 & 0.078$\pm$0.049 & 0.096$\pm$0.091 \\\hline
    $\lambda$ ($\mu$m)& 804$\pm$62 & 1043$\pm$261 & 400$\pm$50 & 628$\pm$69 \\
    $\lambda/W$ & 8.04$\pm$0.62 & 10.43$\pm$2.61 & 4.00$\pm$0.50 & 6.28$\pm$0.69 \\
    $L/W$ & 3.55$\pm$1.11 & 3.51$\pm$1.47 & 5.08$\pm$2.00 & 5.39$\pm$2.17\\
    $\lambda/L$ & 2.26$\pm$0.88 & 2.97$\pm$1.99 & 0.79$\pm$0.41 & 1.17$\pm$0.60\\
    $\lambda_c/W$ & 3.51& 3.32& 3.38& 3.59\\
    $\lambda_c/L$ & 0.99$\pm$0.31& 0.95$\pm$0.40& 0.85$\pm$0.26& 0.66$\pm$0.27\\\hline
    $A$ ($\mu$m)& 70$\pm$15 & 36$\pm$8 & 49$\pm$15 & 73$\pm$19 \\
    $\Upsilon$ & 0.97$\pm$0.40& 0.63$\pm$0.32& 0.94$\pm$0.54& 0.89$\pm$0.45\\
    $\omega_{fit}$ (s$^{-1}$) & 2.8$\times$10$^{5}$& 5.3$\times$10$^{4}$& 2.7$\times$10$^{5}$& 1.2$\times$10$^{5}$\\
    $(D/W)_{fit}$ & 8.6$\times$10$^{-3}$& 1.8$\times$10$^{-2}$ & 9.9$\times$10$^{-3}$& 1.1$\times$10$^{-2}$\\
\end{tabular}
\end{ruledtabular}
\end{table}

\begin{figure}
    \centering
    \includegraphics[width=0.8\linewidth]{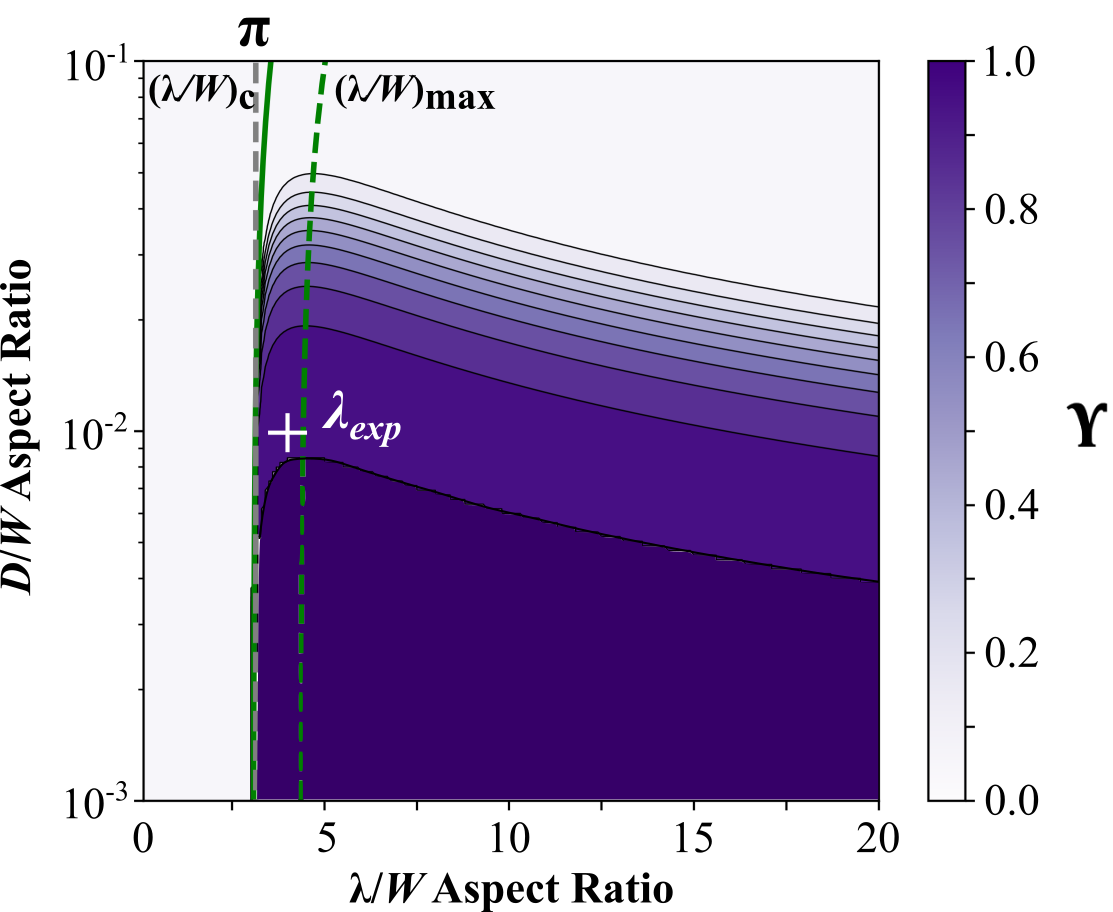}
    \caption{Model prediction for $\Upsilon$ in the thin-wall condition using the experimentally measured solidification rate and melt pool geometry for the 300W \& 500mm/s case, assuming a $\eta/(D+h)=5\%$.}
    \label{fig:thin_wall-prediction}
\end{figure}

Several important features arise from this analysis. In the regime of $D/W$ approaching zero, balling is large because the fluid is highly unstable and solidification is slow without side-walls to conduct heat away. Moreover, the fluid instability growth rate is very sensitive to the solidification front's curvature in this limit, dramatically influencing the balling fraction. The sensitivity of balling fraction to the depth-to-width aspect ratio is shown in Fig. \ref{fig:thin_wall-prediction} for Case C, which is most representative of the fluid instability case, as will be discussed in the following paragraph. As $D/W$ decreases for a fixed melt pool thickness $h$, the threshold wavelength for the Rivulet instability approaches a value similar to that predicted by the Plateau-Rayleigh instability model, and the growth rate of the fluid instability increases because the melt pool curvature is decreasing and thus offering less stabilization. It should be noted however, that as $D/W\to0$, $\beta\to\pi/2$ from Eq. \ref{eq:beta}, and $1-\mathfrak{r}\to\infty$ from Eq. \ref{eq:roundedness-explicit} indicating the dimensionless fluid flow formalism adapted to AM conventions is diverging and the Rivulet model is breaking down because the rivulet itself does not exist. For this reason, our Rivulet model captures the appropriate behavior but may not behave physically for shallow melt pools in the undermelting / lack-of-fusion regime, as currently constructed.

\begin{figure*}
    \centering
    \includegraphics[width=0.8\linewidth]{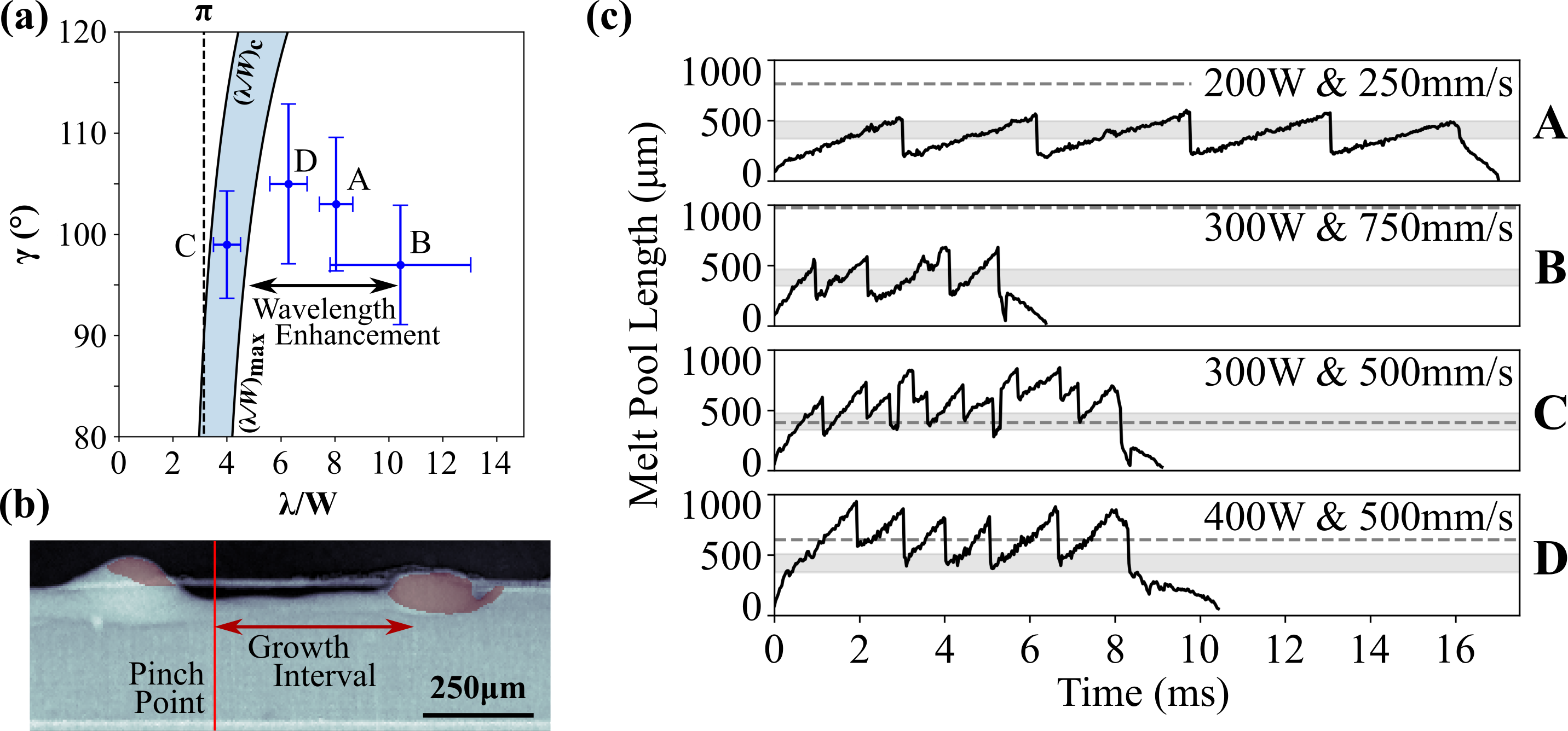}
    \caption{(a) Range of expected fluid instability wavelength versus fluid angle $\gamma$, with experimentally measured values. (b) Example radiograph showing fluid instability delay during melt pool growth after fragmentation from the 300W \& 750mm/s melt track. Red overlays are the melt pool segmentations. (c) Shows the melt pool lengths over time for each case with grey region highlighting the wavelengths expected from Rivulet instability theory and the dashed line showing the observed wavelength.}
    \label{fig:wavelength-shift}
\end{figure*}

Most importantly, comparison to experimental results reveals a discrepancy between the measured wavelengths and the wavelengths predicted by the Rivulet instability theory, which are independent of $D/W$ for very shallow melt pools. Fig. \ref{fig:wavelength-shift}(a) shows that except for Case C, the experimentally measured wavelengths are much larger than the maximum wavelength anticipated from the fastest growing instability wavelength. The reason for this wavelength enhancement is revealed from the radiographs and shown in Fig. \ref{fig:wavelength-shift}(b). After the melt pool necks and fragments, part of the melt pool remains isolated and solidifies into the ball, as our model presumes. But the active melt pool trailing the laser is now significantly shorter than the critical unstable wavelength and does not immediately experience the Rivulet instability. Instead, this active melt pool slowly grows in length until it is once again long enough to support a fluid instability, creating a saw-tooth profile in $L$ vs $t$ shown in Fig. \ref{fig:wavelength-shift}(c). The laser scanning and melt pool motion during this growth interval acts to ``insert" extra track length between balls, effectively increasing the wavelength of the observed balling beyond what would be predicted by the fluid instability.

\begin{figure}[h!]
    \centering
    \includegraphics[width=\linewidth]{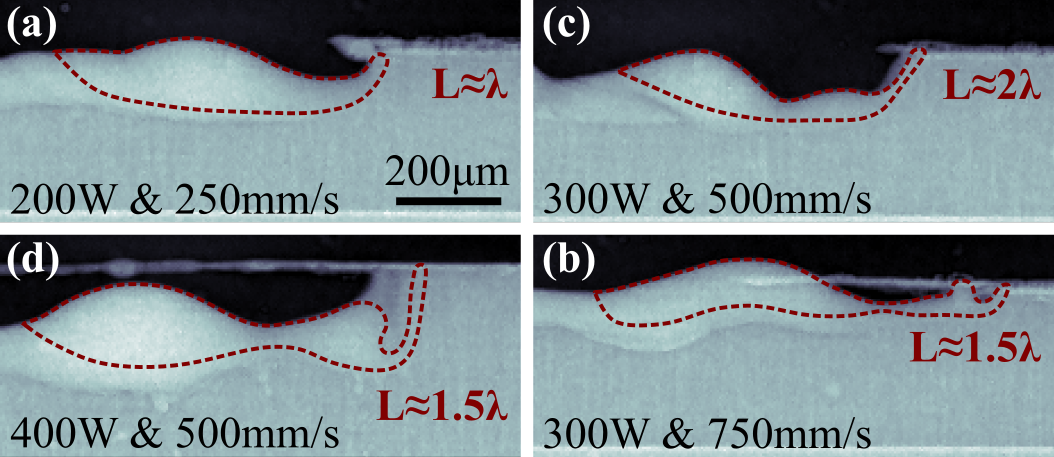}
    \caption{(a) - (d) Representative frames of the radiography with the melt pool outlined in red dashed lines. In all cases, the melt pool surface has the shape of a sinusoid with varying numbers of wavelengths.}
    \label{fig:mode-shapes}
\end{figure}

Recall our assertion that $\lambda\neq L$ in the most general case, despite the AM community's assumption that wavelength is equal to melt pool length. From Fig. \ref{fig:wavelength-shift}(c), the saw-tooth profile of melt pool length over time is seen to often drop below the threshold wavelength $\lambda_c$ (for which $\omega=0$) and grows to significantly larger than the most unstable wavelength which would be expected to be the longest observed wavelength $\lambda_{max}$ without the wavelength enhancement just discussed. The disparity between $\lambda$ and $L$ is clearly seen in our data and may help to explain variation in literature wavelengths based on traditional Plateau-Rayleigh instability arguments \cite{Yadroitsev2010, Gratzke1992, Francis, Leung2022, Bhatt2023}. Due to melt pool motion during its growth phase to unstable lengths, $\lambda>\lambda_{max}$ for reasons which cannot be predicted by either Plateau-Rayleigh instability theory or our use of the Rivulet instability theory. Even without wavelength enhancement, the melt pool length can greatly exceed the predicted $\lambda_{max}$. Fig. \ref{fig:mode-shapes} shows examples from the radiography where the top surface of the melt pool can vary between 1--2 wavelengths. At time steps where the melt pool is in its growth phase, only half a wavelength may be present in the fluid surface. It is important to note however, that the wavelength in the fluid surface is more representative of the fluid instability wavelength than the solidified wavelength. In the Supplemental text, we propose an analysis to predict the initial perturbation amplitude which predicts that an arbitrary fluid perturbation might excite many wavelengths of fluid instabilities which are possible in a given melt pool length. We hypothesize that the fastest growing of these excited wavelengths will be the fluid instability wavelength experimentally observed, offset by the wavelength enhancement due to melt pool drift.

In this section, we have presented a treatment of how our Rivulet+Solidification model can be adapted to the thin-wall case when the laser melts through the entire sample thickness, discussing both the trend of the balling fraction and how the Rivulet theory used here breaks down when the melt pool has no depth into the substrate. We further discussed a mechanism by which the experimentally observed wavelength can be significantly larger than the predicted fluid instability wavelengths due to laser scanning during the melt pool growth phase after a balling event. This mechanism, as well as radiography of the melt pool surface assert that $\lambda\neq L$ in general.

\section{\label{sec:implications}Implications of our Model}

\begin{figure*}
    \centering
    \includegraphics[width=0.7\linewidth]{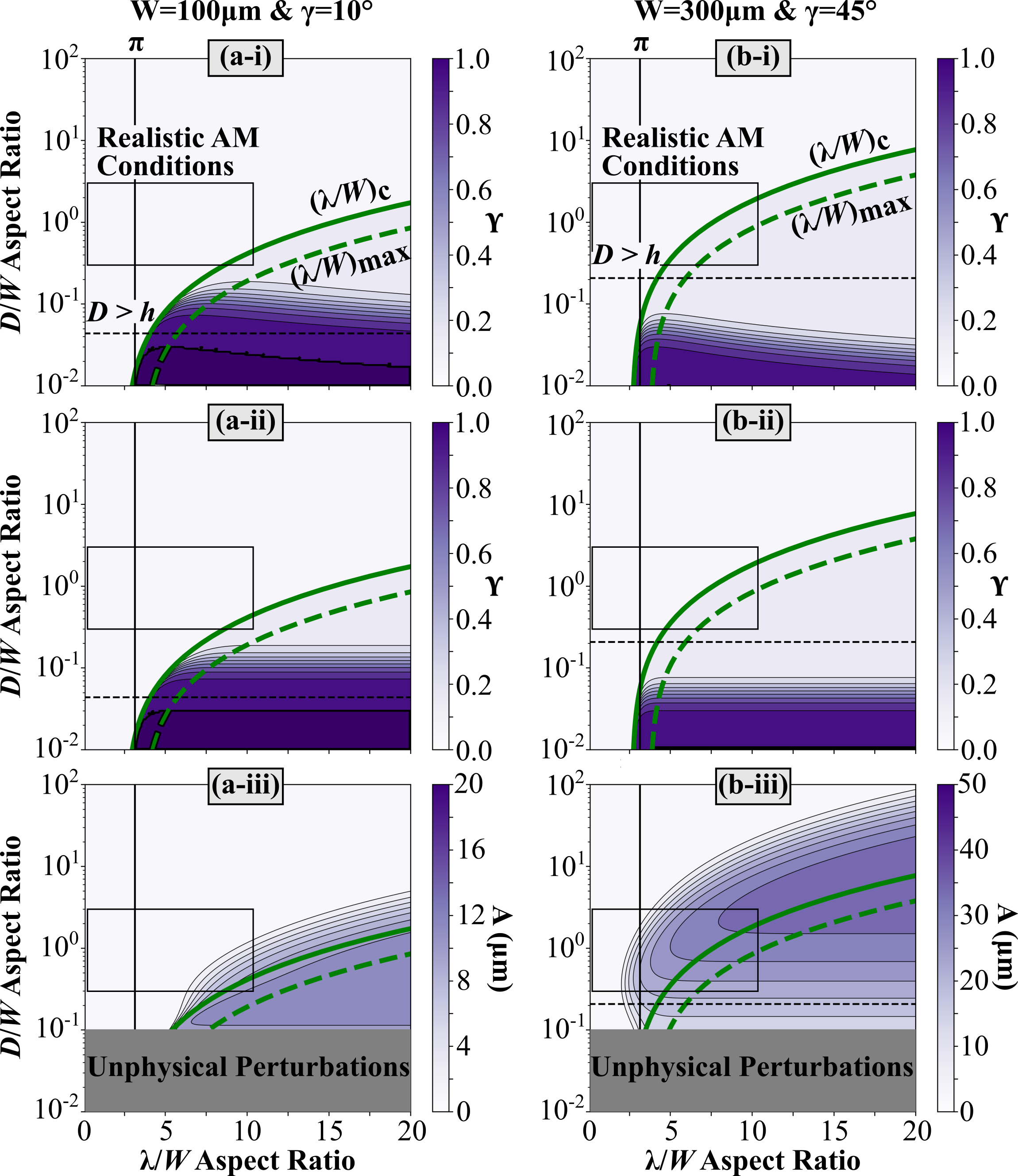}
    \caption{Comparing balling fraction and absolute amplitude for two melt pool geometries (a) and (b). For the instantaneous $\lambda/L$ (i) and largest accessible wavelength (ii) plots, the initial perturbation is taken to be $\eta_0/(D+h)=10\%$. The absolute balling amplitude (iii) plots have a constant initial perturbation amplitude fixed at $\eta_0/W=5\%$.}
    \label{fig:upsilon-progression}
\end{figure*}

In this section, we discuss the implications of our model beyond the experiments done in this work to illustrate how they compare to the more commonly used sample conditions relevant to laser welding and AM, for which our model was developed. Our discussion in this section centers around the more general cases shown in Fig.~\ref{fig:upsilon-progression} which describes an analysis progressing from the balling fraction at a given wavelength to the absolute amplitude observed at that wavelength, relevant to avoiding defect formation. We carry out this analysis for two melt pool dimensions: the (a) series is calculated for $W=100\mu$m wide melt pools with fluid surface angle $\gamma=10^\circ$, while the (b) series is calculated for $W=300\mu$m and $\gamma=45^\circ$. 

The Rivulet instability model we have presented defines a fluid instability growth rate at any given wavelength for a given melt pool geometry, which is coupled to solidification by our combined model to give $\Upsilon$ shown in Fig. \ref{fig:upsilon-progression}(a-i) and (b-i). As the melt pool becomes more shallow, the substrate offers less stabilization and the balling fraction increases. The solid green curve in Fig. \ref{fig:upsilon-progression} describes the threshold wavelength $\lambda_c$ for which $\omega=0$ while the dashed green curve depicts the wavelength of the fastest growing wavelength $\lambda_{max}$, both of which are dependent on the melt pool depth-to-width aspect ratio. However, all fluid wavelengths that are longer than the fastest growing wavelength grow more slowly and $\lambda_{max}$ should dominate over them. Fig. \ref{fig:upsilon-progression}(a-ii) and (b-ii) show how the balling fraction for $\lambda>\lambda_{max}$ is given by the value at $\lambda_{max}$; these would be the experimentally observed fluid balling fractions expected from Rivulet instabilities, accounting for the competition between different wavelengths. Wavelengths shorter than the fastest growing wavelength are presumed to be the only wavelengths that can exist within the melt pool at that condition, and are thus the fastest growing available wavelength.

In Fig. \ref{fig:upsilon-progression}(a-i) and (b-i), the balling fraction is strongly influenced by the fluid material above the substrate surface. The horizontal dashed lines depict where $D=h$ and separate the regimes where the melt pool fluid instability behavior is dominated by the fluid above or below the substrate surface.  The melt pool height above the substrate $h$ is independent of the $D/W$ aspect ratio and fixed by the scale parameter $W$ and $\gamma$ only. Larger $h$ increases the effective solidification time and the time available for the fluid instability to contribute to balling. However, as the melt pool becomes deeper with respect to its width, the fluid volume is predominantly below the substrate surface and the Rivulet instability is less strongly influenced by the $\gamma$ parameter.

As we discussed in Section \ref{sec:adapting-rivulet-to-AM}, the shape of the balling contour helps to explain the negative relationship between the laser scan velocity and the balling wavelength observed in the literature. As the laser scan velocity increases, the depth-to-width aspect ratio decreases which serves to destabilize the melt pool and lower the critical threshold for balling, reducing the onset wavelength of the balling.

While the balling fraction is generally highest for shallow melt pools, the absolute amplitude of balling scales with the melt pool depth. Very shallow melt pools with complete balling have low balling amplitude and would most likely be construed as lack-of-fusion or undermelting in experiments. So, while the balling fraction $\Upsilon$ is an important parameter for explaining the overall behavior, an experimentalist is most concerned with the absolute amplitude of balling, $A=\Upsilon\cdot(D+h)$, which results in defects at the part scale. For this reason, shallower melt pools have a larger balling fraction but a smaller absolute amplitude while deeper melt pools can experience lower balling fractions but larger absolute amplitudes. This competition creates the ``nose" feature observed in Fig. \ref{fig:upsilon-progression}(a-iii) and (b-iii). Simultaneously deeper and longer melt pools would be expected to have larger absolute balling amplitudes.

It should also be noted that there is a non-zero balling fraction predicted by our theory even in the ``stable" region of the geometry, above the solid green line in Fig. \ref{fig:upsilon-progression}(a-iii) and (b-iii). The gradual transition from non-balling to balling is accounted for by our model and seen in the gradient above the green $\lambda_c$ line. In the ``stable" region, the fluid instability growth rate is negative, indicating that with sufficient time a fluid perturbation would exponentially decay away to give no balling amplitude. In this region near the critical threshold, the magnitude of the growth rate is small relative to the solidification rate and there is insufficient time for a perturbation to decay fully before solidification; the result is that the hills are ``frozen in" for the final printed sample. This nuance can help explain why balling can be observed even under conditions that would normally be considered stable.

In Section \ref{sec:theory-to-experiments}, we discussed how experiments show $\lambda\neq L$ as a result of the melt pool displacement between balling events and how the surface profile of the melt pool can contain more or less than one period of oscillation prior to fragmentation.

Beyond wavelength alone, the growth rates of the instability have also been observed to change based on edge effects like those observed in AM. The finite length of the melt pool can initiate ``pearling" or an enhanced rate of breakup from the track ends where balling may already be initiating \cite{Diez2009, Gonzlez2007}. This pearling is known to have a faster timescale than perturbation growth in an infinite-length melt pool, which would contribute to greater balling fractions. Pearling can also cause small deviations away from the wavelengths predicted by rivulet analysis. Due to the boundary conditions of a melt pool, pearling is always present, but we do not account for this effect in our rivulet-based analysis. Finally, we comment that for highly viscous liquids (based on the Ohnesorge number, $Oh=\mu/\sqrt{\rho\sigma R}\gtrsim 0.2$ for a jet) or at the end of the solidification process, the exponential growth of the instability from the rivulet models breaks down \cite{Dumouchel2015, Clasen2011}. For systems considering the late times of the balling process, we note that further consideration of the sub-exponential kinetics may be worth accounting for.

In general, the fastest growing wavelength supported by the melt pool will result in the greatest balling amplitude due to fluid instabilities. Even neglecting more complex physics like fluid flows within the melt pool, the above considerations offer context on how the complexity of balling makes the convenient assumption that $\lambda=L$ break down. 
In the most general case, any wavelength could be supported which adheres to conservation of volume and the boundary conditions that define the melt-pool track.

\section{\label{sec:limitations}Theory Limitations and Advanced Considerations}

In this section, we note the limitations of our present model and discuss how the framework we present to relate instability theory to solidification may be adapted to more accurately describe the AM conditions.

The Plateau-Rayleigh instability model commonly used by the AM community assumes the fluid instability is independent of solidification (or the presence of any surfaces), negligible gravitational effects, no pre-existing fluid flows within the melt pool, no turbulent flow, no powder effects, and simultaneously an infinite-length melt pool and $\lambda=L$; see Section \ref{sec:instability-theory}. Rivulet instability theory relaxes the constraint that the fluid exists in isolation from all other surfaces and is used to account for the stabilizing effect of the substrate on the melt pool. We took a Rivulet model for general geometry, adapting it to AM conditions and combined it with solidification in a 1D model for balling based on melt pool fragmentation. Melt pool fragmentation assumes slow solidification relative to fluid instability growth rate, or alternatively long melt pools for which the liquid-gas interface intersects the solid-liquid interface discontinuously inside the melt pool rather than continuously from the rear of the melt pool. We also consider the more general case that $\lambda\neq L$. Together, our model of Rivulet instability competing with solidification gives a gradual transition between non-balling and balling conditions. But all other incorrect assumptions regarding the fluid, including the absence of internal fluid flows and powder remain. The mathematical formalism used to account for the roundedness of the melt pool also breaks down when the melt pool's $D/W$ aspect ratio becomes very small.

Furthermore, we note that this work explicitly predicts balling for all combinations of melt pool geometry in 3D space, $L$, $W$, $D$, $\gamma$, initial perturbation $\eta_0$ and solidification rate $R$, without considering how these parameters relate to each other or process parameters native to AM ($P$, $v$, etc.). Substituting computational or analytical models such as the Rosenthal equation can be used to relate the balling model presented in this paper to AM process parameters, but this is beyond the scope of this present work.

Our formalism to account for the competition between solidification and fluid instabilities makes a critical assumption that the competition arises from solidification racing towards the valley of the pinching point. This intrinsically assumes that the melt pool is long and solidification from the tail of the melt pool is not governing the balling phenomena. In practice, such a condition is not always the case. Further, when $\lambda>L$, that valley may not necessarily be present, causing the model to break down. An important aspect of this compound model we have presented is the importance of the two separate aspect ratios ($\lambda/W$ and $D/W$) as well as the amount of powder incorporation and solidification velocity, neither of which is typically included in typical predictions of balling onset presented in the literature.

Using the formalism provided in the current work, we anticipate that future studies will now be able to refine the models describing the AM phenomena of the melt pool and solidification to increase their accuracy. Several improvements of this rivulet-based model could be envisioned to account for the finite melt pool length. A fluid mechanics re-derivation of a fluid instability in an elliptical bowl with differing aspect ratios would more physically describe a melt pool than an infinite-length rivulet in a rounded wedge. The application of pearling-based theories might also be sufficient to account for the melt pool's finite length and the increased fluid instability growth rates when the tail of the melt pool is already experiencing balling due to the previous, solidified hill. A more detailed analysis of the initial perturbation can also be derived which imposes time-varying boundary conditions as the rear of the melt pool solidifies up and down successive balls.

Under process conditions typically encountered in LPBF, balling is often the result of vapor depression or fluid-driven mechanisms rather than fluid instabilities. Our results from X-ray radiography experiments, shown in Figs.~\ref{fig:TimeSeries} and~\ref{fig:waterfall}, provide direct evidence of fluid-driven balling. In these experiments (supplementary videos), we observe fluid from the melting front being driven to the tail of the melt pool. Upon contact with the solid material at the melt pool tail, the fluid starts to accumulate, forming a hill. Our data shows that the hill grows as it is continuously fed with fluid through a liquid channel, as demonstrated in Fig.~\ref{fig:TimeSeries}. This provides direct visual confirmation for experimental and simulation work in the literature \cite{Berger2011, Otto_2016, Wu_2017}. Although fluid-driven balling mechanisms are typically expected to occur at higher scan speeds, we observe them at much lower scan speeds due to the use of thin substrates, which, upon melting, lack side walls. However, fluid-driven balling requires further characterization and quantification beyond the scope of the present work.

\section{\label{sec:summary}Summary}

The Rivulet instability model was adapted to describe balling in AM in the absence of fluid flows. The modified Rivulet model enables the development of a stability diagram as a function of melt pool geometry (depth and width) and any wavelength while accounting for the stabilizing effect provided by the substrate's support below and to the sides of the melt pool. The basic finding is that as the height of the melt pool above the substrate increases, it becomes more unstable as more fluid destabilizes. In contrast, deeper melt pools relative to their width are more stable than shallow ones. This is because the instability growth rate is higher for melt pools with smaller depth-to-width aspect ratios, which are less stabilized by the substrate. Additionally, the instability decay rate is greater for shorter, deeper melt pools, indicating ``stable" conditions in classical balling arguments.

The competition between fluid instabilities and solidification was examined using the average vertical solidification rate and average fluid instability growth rate from the X-ray radiography data. The balling amplitude was predicted to increase with larger melt pool sizes at a constant solidification rate. This is because the melt pool remained in the fluid state for longer, allowing more time for instabilities to grow. Conversely, larger fluid instability growth rates also resulted in greater balling amplitudes. Examination of experimental data revealed that the thin substrate limit favors large and unstable growth rates for fluid instability. When comparing the experimental data with the Rivulet-solidification model, we observe that the model breaks down for vanishingly shallow melt pools. This occurs because fluid instability increases with decreasing melt pool curvature.

Finally, we note that this analytical work provides a description of fluid instabilities in the melt pool and accounts for their competition with the solidification rate as a function of melt pool geometry. These analytical results are calculated for a range of melt pool dimension combinations. The utility of the predictions can be improved by incorporating experimentally measured melt pool dimensions and solidification rates, coupling these parameters to process parameters, and accounting for the interdependence of fluid instability growth rates and solidification rates.

\section*{Supplementary Material}
Further information about the synchrotron image processing and measurement methodology, as well as an estimation of the initial fluid perturbation amplitude is provided in the supplementary material.

\begin{acknowledgments}
We thank Lichao Fang and Adrian Lew for useful discussions. This work was supported by Lawrence Livermore National Laboratory under subcontract No.~B654297, and performed under the auspices of the U.S. Department of Energy by LLNL under Contract No.~DE-AC52-07NA27344. CLAL and PDL are grateful for the support from the UKRI—EPSRC, Grants Numbered EP/W006774/1, EP/P006566/1, EP/W003333/1, and EP/V061798/1. PDL is funded by the support from a Royal Academy of Engineering Chair in Emerging Technologies (CiET1819/10); CLAL is funded in part by EP/W037483/1 and IPG Photonics/ Royal Academy of Engineering Senior Research Fellowship in SEARCH (ref: RCSRF2324-18-71). This research used resources of the European Synchrotron Radiation Facility (ESRF) in Beamline ID19 (ME-1573). Many thanks to team members from the Materials, Structure and Manufacturing group at Harwell (MSM@H) for their assistance in preparation for and during the beamtime.
\end{acknowledgments}

\section*{Conflict of Interest}
The authors declare no competing interests.

\section*{Author Contributions}
\textbf{Zane Taylor:} Conceptualization (lead); Methodology (lead); Formal analysis (equal); Investigation (equal); Data curation (equal); Software (lead); Validation (lead); Visualization (lead); Writing - original draft (equal); Writing - review and editing (equal). \textbf{Tharun Reddy:} Conceptualization (lead); Formal analysis (equal); Investigation (equal); Data curation (equal); Writing - original draft (equal); Writing - review and editing (equal). \textbf{Maureen Fitzpatrick:} Investigation (supporting). \textbf{Kwan Kim:} Investigation (supporting). \textbf{Wei Li:} Investigation (supporting). \textbf{Chu Lun Alex Leung:} Investigation (lead); Resources (equal). \textbf{Peter D. Lee:} Investigation (supporting); Resources (equal). \textbf{Kaila M. Bertsch:} Supervision (supporting). \textbf{Leora Dresselhaus-Marais:} Conceptualization (supporting); Supervision (lead); Project administration (lead); Funding acquisition (lead); Writing - review and editing (equal). 

\section*{Data Availability Statement}

The data that support the findings of this study are available upon reasonable request from the corresponding author. Raw data were generated at European Synchrotron Radiation Facility.

\appendix
\section{Variables Used in This Work}

\begin{longtable}{cl}
        \hline\hline
        Variable & Parameter \\\hline
        $a_0$ & radius of rivulet wedge \\
        $a_1$ & distance from origin of liquid-air interface to \\&wedge corner \\
        $a$ & approximately the curvature of the wedge corner \\
        $a_c$ & radius of curvature of liquid-air interface \\
        $A$ & balling amplitude \\
        $B$ & geometric rivulet factor \\
        $C$ & geometric rivulet factor \\
        $D$ & melt pool depth below substrate surface \\
        $E$ & geometric rivulet factor \\
        $h$ & melt pool height above substrate surface \\
        $I_k$ & modified Bessel functions of the $k$-th kind \\
        $k$ & wavenumber of fluid instability \\
        $\widetilde{k}$ & dimensionless wavenumber of fluid instability \\
        $K$ & geometric rivulet factor \\
        $L$ & melt pool length \\
        $P$ & laser power \\
        $Q$ & dimensionless volumetic flow of rivulet \\
        $\mathfrak{r}$ & roundedness of melt pool in rivulet wedge \\
        $R$ & vertical solidification rate, later the \textit{average} vertical\\& solidification rate \\
        $R_t$ & vertical solidification rate at the end of solidification \\
        $R_0$ & initial fluid radius of Plateau-Rayleigh jet \\
        $t$ & time \\
        $t_0$ & rivulet time constant \\
        $\widetilde{t}_0$ & normalized rivulet time constant \\
        $v$ & laser scan velocity \\
        $W$ & melt pool width \\
        $y$ & position in axis parallel to laser \\
        $x$ & position in axis parallel to melt pool travel \\
        $z_0$ & wavenumber normalization factor \\
        $\beta$ & angle of melt pool wall with substrate surface normal \\
        $\gamma$ & contact angle of melt pool surface with the\\&substrate surface \\
        $\eta$ & fluid perturbation profile \\
        $\eta_0$ & initial fluid perturbation amplitude \\
        $\theta$ & wetting angle \\
        $\lambda$ & wavelength of the fluid instability \\
        $\lambda_{c}$ & critical wavelength at which $\omega=0$ \\
        $\lambda_{max}$ & fastest growing fluid instability wavelength \\
        $\mu$ & viscosity \\
        $\xi$ & distance along $x$ behind the laser \\
        $\rho$ & density \\
        $\sigma$ & surface tension \\
        $\tau$ & time at which solidification front meets the\\& melt pool surface \\
        $\Upsilon$ & dimensionless balling fraction \\
        $\phi$ & phase of sinusoidal instability \\
        $\omega$ & imaginary component of fluid instability growth rate \\
        $\widetilde{\omega}$ & dimensionless fluid instability growth rate \\
        $\omega_{max}$ & $\omega$ at $\lambda_{max}$ \\
        $\Omega$ & fluid instability growth rate as a complex number \\\hline\hline
\end{longtable}

\nocite{*}
\bibliography{references_USED}

\end{document}


\begin{frontmatter}

\title{Supporting Information for: Analytical model for balling defects in laser melting using rivulet theory and solidification}
 
\author[1,2]{Zane Taylor}
\author[1,2]{Tharun Reddy}
\author[3]{Maureen Fitzpatrick}
\author[4,5]{Kwan Kim}
\author[4,5]{Wei Li}
\author[4,5]{Chu Lun Alex Leung}
\author[4,5]{Peter D. Lee}
\author[6]{Kaila M. Bertsch}
\author[1,2]{Leora Dresselhaus-Marais\corref{cor1}}
\ead{leoradm@stanford.edu}
\cortext[cor1]{Corresponding author: 
}
\affiliation[1]{organization={Department of Materials Science and Engineering, Stanford University},
            city={Stanford},
            state={CA},
            postcode={94305}, 
            country={USA}}
\affiliation[2]{organization={SLAC National Accelerator Laboratory},
            city={Stanford},
            state={CA},
            postcode={94025}, 
            country={USA}}

\affiliation[3]{organization={European Synchrotron Radiation Facility},
            city={Grenoble},
            postcode={38000}, 
            country={France}}
            
\affiliation[4]{organization={Department of Mechanical Engineering, University College London},
            city={London},
            state={WC1E 7JE},
            country={UK}}
            
\affiliation[5]{organization={Research Complex at Harwell, Rutherford Appleton Laboratory},
            city={Oxfordshire},
            state={OX11 0FA},
            country={UK}}

\affiliation[6]{organization={Lawrence Livermore National Laboratory},
            city={Livermore},
            state={CA},
            postcode={94550}, 
            country={USA}}
            
\end{frontmatter}
 
\makeatletter 
\renewcommand{\thefigure}{S\@arabic\c@figure}
\renewcommand{\thetable}{S\@arabic\c@table}
\makeatother

\section{Methods}
\subsection{Conversion of intensity maps to thickness maps}
X-rays are attenuated as they pass through a medium, decreasing X-ray intensity as they penetrate deeper into the medium. The attenuation is influenced by (i) the atomic number of the elements in the medium, characterized by the characteristic mass absorption coefficient of the medium, and (ii) the medium's thickness. The intensity of the incident X-ray beam (\( I_{\text{o}} \)) passing through the sampling medium (100 µm of tantalum in our experiment) decreases exponentially with the distance traveled. The transmitted X-ray intensity (\( I \)) is measured after penetration through the sample, and the relationship between \( I_{\text{o}} \) and \( I \) is described by Beer-Lambert's law,
\begin{equation}
    I = I_0 \exp\left( -\frac{\mu}{\rho} \cdot t \cdot \rho \right)
    \label{eq:beer-lambert}
\end{equation}

where ${\mu}$/${\rho}$ is the mass attenuation coefficient, $\rho$ is the material density, and $t$ is the thickness of sampling material. In this study, the \( I \)  and \( I_{\text{o}} \) values are measured from the raw unprocessed X-ray radiograph pixels with 16-bit grayscale values. Given the synchrotron polychromatic X-ray beam with a peak at $\sim$30 keV and the undulator producing higher energy harmonics up to 90 keV, an average transmitted X-ray energy value can be estimated using equation \ref{eq:beer-lambert}. The \( I \)  and \( I_{\text{o}} \) values, along with $t$ = 100 µm and $\rho$ = 16.6 $g/cm^2$ were used to calculate ${\mu}$/${\rho}$. Using the calculated ${\mu}$/${\rho}$, the average X-ray energy was determined with data available in the NIST database \cite{NISTXRay}, where ${\mu}$/${\rho}$ is listed as a function of X-ray energy. The average ${\mu}$/${\rho}$ through 100 µm of tantalum sample is $\sim$10.793 $cm^2/g$ and corresponds to 40 keV in the NIST database. Using the ${\mu}$/${\rho}$ value at 40 keV, variations in \( I \) can be used to estimate the change in thickness of the sample as shown in Figs.~2(a) and 9. This method is utilized to convert the intensity maps into thickness maps. The thickness measurement provides the through-plane dimension, while the effective pixel size of 4.126 µm x 4.126 µm defines in-plane dimensions.

\subsection{Solidification and free surface rate measurement}
\label{sec:rates}
The balling mechanism is driven by the competition between the solidification rate and the free surface advancement rate, which is influenced by fluid instabilities, with both processes progressing in opposite directions. Therefore, accurately measuring their respective timescales is essential. These measurements are obtained from the radiography images, where a single 16-bit grayscale radiograph image can be represented as a 2D array with intensity values ranging from 0 to 65,536. A specific column from this array, corresponding to the region of interest, can be extracted for analysis. By tracking the intensity values of this column across all images in a time sequence, temporal variations can be observed. These variations are visualized using a waterfall plot, which reveals changes in intensity along a particular column throughout the melting and solidification cycles. The slope of the profile in the waterfall plots provides the solidification rate $R_t$ and free surface advancement rate $R_f$ at the end of solidification as shown in Fig.~2(b). 

\subsection{Measuring melt pool dimensions}
The melt pools were hand segmented in each frame of the videos for our four cases. Since the aim of this study is primarily theoretical, further automation of melt pool segmentation was not performed. From these segmentations, most of our automated measurements of melt pool dimensions were performed.

Melt pool length $L$ was measured from the segmented images as the horizontal length of the right-most continuous melt pool. The thickness of the melt pool $D+h\approx h$ was measured as the average distance between the top and bottom surface of the segmented melt pool. If two melt pools were in the same frame, due to balling and fragmentation, only the melt pool being fed by the laser was evaluated.

Balling wavelength $\lambda$ was taken as the horizontal distance between the peak locations identified from the final frame of each video. Balling amplitude $A$ was calculated as half the average of the peak-to-valley distance. Using thickness plots calculated using Beer-Lambert's law, the angle of the solidified surface in each column of pixels was calculated, and the average of the angles which were greater than $\pi/2$ was reported as $\gamma$. All of $\lambda$, $A$, and $\gamma$ were measured using the final frames of each video.

In all cases, the standard deviation of measurements was calculated. This standard deviation captures the variance of experimentally measured values, but not the errors in the measurements. More involved error estimation was not performed because the aim of this study is largely theoretical, and no ground truth is available.

\subsection{Variance Propagation}
In several cases, we report the standard deviation of a fraction of terms with standard deviations. We evaluate these in the following way.

For $L=\bar{L}\pm \sigma_L$ and $\lambda=\bar{\lambda}\pm\sigma_\lambda$,
\begin{align}
    \frac{1}{L}& = \frac{1}{\bar{L}}\pm\frac{\sigma_L}{\bar{L}^2}\\
    \frac{\lambda}{L}&=\frac{\bar{\lambda}}{\bar{L}}\pm\bigg(\frac{\sigma_\lambda}{\bar{L}}+\frac{\bar{\lambda}\sigma_L}{\bar{L}^2}\bigg)
\end{align}

\section{Fourier Transform Estimation of $\eta_0$}
Consider the melt pool surface at $t_0$ is defined by the function $y(x)$. From the linear instability theory upon which the Plateau-Rayleigh and Rivulet models are based, every wavelength can be treated as independent and non-interacting. The Fourier transform of $y$ describes $y$ in terms of its spatial frequency (or wavelength) components.

Thus,
\begin{equation}
    \eta(k) = \mathcal{F}\{y(x)\}
\end{equation}
and the initial perturbation using in the calculation of the fluid instability growth rates can be calculated from the initial melt pool surface profile for any given wavelength.

Note, there is a complication that the melt pool surface profile is evolving with time due to laser translation, fluid flow, and instability growth, so the initial perturbation amplitude is also changing with time in a non-trivial way. The complication of time and continuity of the surface profile is left for a later work.

If we impose the boundary conditions that $y(x)=0$ for $x=0,L$ at the beginning and end of the melt pool. The Fourier transform reduces to a sine transform which must go to zero at the boundaries. Further, this sine transform now has the constraint that
\begin{equation}
    \lambda = \frac{2L}{n}
\end{equation}
where $n\in\mathbb{Z}_+$. This constraint ensures that the total length contains only an integer number of half-wavelengths, which is the same as saying $y(x)$ goes to zero at the edges. This result is equivalent to particle-in-a-box or vibrating string models as a result of the boundary conditions.

The most important thing to note from this simple result is that $\lambda=2L$ is a valid solution, as are other solutions for which $\lambda\neq L$. The fastest growing wavelength that can be supported by the boundary conditions and constraints of the melt pool surface will define the periodicity of the fluid instability. A second important note is that if $\lambda=2L$, or $n=1$, then the fluid instability's surface is just a singular hump. In such a case there is not necessarily a valley in which ``pinching" and melt pool fragmentation by the mechanism discussed in this paper can occur. This argument may support why $\lambda_c/L$ is of order unity in our experimental results, even though the two need not be equivalent.

Imposing the boundary conditions we have here is not generally correct, but makes our point that $\lambda$ and $L$ can be related and that an infinite number of wavelengths are theoretically possible. More realistic boundary conditions are $y(0)=0$ at the melt pool's leading edge, where the laser is, and $y(L)$ is a free parameter (as it often is in reality). Another boundary condition is necessary in that case to constrain the valid values of $\lambda$ as a function of $L$, but the boundary condition at the end of the melt pool is complex and not well-enough defined for presentation here.

\bibliographystyle{elsarticle-num-names} 
\bibliography{references_SI}